\documentclass[traditabstract]{aa} 

\usepackage{graphicx}
\usepackage{color}
\usepackage{url}
\usepackage{amssymb}
\usepackage{amsmath}
\usepackage{txfonts}
\usepackage{multirow}
\usepackage{lscape}
\usepackage[flushleft]{threeparttable}
\usepackage{float}
\usepackage{natbib}
\usepackage{stmaryrd}
\usepackage{supertabular}
\usepackage{rotating}
\usepackage{longtable}
\bibpunct{(}{)}{;}{a}{}{,} 

\makeatletter
\def\hlinewd#1{%
\noalign{\ifnum0=`}\fi\hrule \@height #1 %
\futurelet\reserved@a\@xhline}
\makeatother

\begin{document}
   \title{SARCS strong lensing galaxy groups: I - optical, weak lensing, and scaling laws
   \thanks{Strong Lensing Legacy Survey SL2S-ARCS}
}
   \author{
          G. Fo\"ex\inst{1}
          \and
          V. Motta\inst{1}
          \and
          M. Limousin\inst{2,3}
          \and
          T. Verdugo\inst{4}
           \and
          A. More\inst{5,6}
          \and
          R. Cabanac\inst{7}
          \and
          R. Gavazzi\inst{8}
          \and
          R.P. Mu\~{n}oz\inst{9}
          }
   \institute{
          Departamento de F\'isica y Astronom\'ia, Universidad de Valpara\'iso, Avda. Gran Breta\~{n}a 1111, Valpara\'iso, Chile; gael.foex@uv.cl
          \and
          Aix Marseille UniversitŽ, CNRS, LAM (Laboratoire d'Astrophysique de Marseille) UMR 7326, 13388, Marseille, France
          \and
          Dark Cosmology Centre, Niels Bohr Institute, University of Copenhagen; 
               Juliane Maries Vej 30, DK-2100 Copenhagen, Denmark
          \and
          Centro de Investigaciones de Astronom\'ia, AP 264, M\'erida 5101-A, Venezuela
          \and
          Kavli Institute for Cosmological Physics, U. of Chicago, 5640 S. Ellis Ave., Chicago IL-60637, USA
          \and
          Kavli IPMU, U. of Tokyo, 5-1-5 Kashiwanoha, Kashiwa, 277-8583, Japan
          \and
          CNRS; Institut de Recherche en Astrophysique et Plan\'etologie;  57 avenue d'Azereix, 65000 Tarbes, France
          \and
          Institut d'Astrophysique de Paris, UMR 7095 CNRS \& Universit\'e Pierre et Marie Curie, 98bis Bd Arago, F-75014 Paris, France
          \and
          Departamento de Astronom\'ia y Astrof\'isica, Pontificia Universidad Cat\'olica de Chile, V. Mackenna 4860, Santiago, Chile
             }

   \date{Received ; accepted }

  \abstract  
   {We present the weak lensing and optical analysis of the SL2S-ARCS (SARCS) sample of strong lens candidates. The sample is based on the Strong Lensing Legacy Survey (SL2S), a systematic search of strong lensing systems in the photometric Canada-France-Hawaii Telescope Legacy Survey (CFHTLS). The SARCS sample focuses on arc-like features and is designed to contain mostly galaxy groups. We briefly present the weak lensing methodology that we use to estimate the mass of the SARCS objects. Among 126 candidates, we obtain a weak lensing detection (at the $1\sigma$ level) for 89 objects with velocity dispersions of the Singular Isothermal Sphere mass model (SIS) ranging from $\sigma_{SIS}\sim350\,\mathrm{km\,s^{-1}}$ to $\sim1000\,\mathrm{km\,s^{-1}}$ with an average value of $\sigma_{SIS}\sim600\,\mathrm{km\,s^{-1}}$, corresponding to a rich galaxy group (or poor cluster). From the galaxies belonging to the bright end of the group's red sequence ($M_{i}<-21$), we derive the optical properties of the SARCS candidates. We obtain typical richnesses of $N\sim5-15$ galaxies and optical luminosities of $L\sim0.5-1.5\times10^{12}\mathrm{L_{\odot}}$ (within a radius of 0.5 Mpc). We use these galaxies to compute luminosity density maps, from which a morphological classification reveals that a large fraction of the sample ($\sim45\%$) are groups with a complex light distribution, either elliptical or multimodal, suggesting that these objects are dynamically young structures. We finally combine the lensing and optical analyses to draw a sample of 80 most secure group candidates, i.e. weak lensing detection and over-density at the lens position in the luminosity map, to remove false detections and galaxy-scale systems from the initial sample. We use this reduced sample to probe the optical scaling relations in combination with a sample of massive galaxy clusters. We detect the expected correlations over the probed range in mass with a typical scatter of $\sim25\%$ in $\sigma_{SIS}$ at a given richness or luminosity, making these scaling laws interesting mass proxies.}

   \keywords{Gravitational lensing: weak -- Cosmology: observations -- Cosmology: dark matter -- Galaxies: groups : general
               }

   \maketitle

%

\section{Introduction}

From a nearly homogenous and uniform matter density field, the Universe has evolved through cosmic time to a complex distribution of filamentary and clumpy structures. The Universe main matter (dark matter) follows a hierarchical model of structure formation and evolution \citep{kaiser86,white91} depicted in great details with numerical simulations, tracing the gravitational growth of dark matter haloes \citep{evrard02,springel05,dolag06}. The so-called cosmic web filling the present Universe has been observationally confirmed by several large spectroscopic surveys \citep{colless01,pimbblet04,pandey06}, revealing the presence of large-scale filaments 'feeding' nodes of massive and rich galaxy clusters (e.g. \citealt{jauzac12}). In the picture of this evolving matter density field, the intermediate-mass range of the galaxy groups play a key role in structure formation as they contain the majority of all galaxies (at least at low redshifts, \citealt{eke04}) and bridge the gap between large massive galaxy clusters and single galaxies.\\
A precise characterization of the total mass contained in groups and clusters of galaxies and its connection to the visible baryonic tracers is one of the most important and yet challenging goal for cosmological and astrophysical purposes. For instance, the group and cluster mass function (number density of object as function of total mass) and its redshift evolution is one of the most powerful cosmological constraint as it is sensitive both to the Universe expansion and the growth rate of structures (e.g. \citealt{white93,haiman01,wang04,rozo09}). To be fully effective, this cosmological probe requires large samples of groups and clusters with precise mass estimates. Several methodologies can provide direct mass measurements, such as the analysis of the X-ray emission of the hot intra cluster medium (ICM), dynamical studies of galaxy velocity dispersions or with the gravitational lensing signal produced on background galaxies. However, all these techniques require high-quality data sets and non-trivial analyses. It is therefore more convenient to make use of baryonic tracers as mass proxies to derive quickly masses for large numbers of groups and clusters. In the simplest model of structure formation involving a purely gravitational collapse of dark matter haloes, groups and clusters form a population of self-similar objects with simple relations between their total mass and other physical quantities \citep{kaiser86}. Numerous works have explored and tried to fully characterize these links between mass and baryonic observables such as the ICM X-ray luminosity, temperature, pressure or entropy from X-ray observations \citep{finoguenov01b,vikhlinin02,ettori04,arnaud05,kotov05,vikhlinin06,hoekstra07,rykoff08b,pratt09,leauthaud10,okabe10a,mahdavi12}, the Compton parameters derived from Sunyaev-Zel'dovich (SZ) observations \citep{mccarthy03,morandi07,bonamente08,marrone09,lancaster11,planck6,planck11}, or the galaxy velocity dispersions, richness and optical luminosity from optical observations \citep{lin03,lin04a,popesso05,becker07,johnston07,popesso07,reyes08,mandelbaum08,rozo09,andreon10,foex12}. However, many observational results have found discrepancies with the theoretical predictions derived from the gravitationally driven model of structure formation (different slope and normalization, break of self-similarity at low mass, large intrinsic scatter, non-standard redshift evolution), thus revealing the combined influence of various non-gravitational physical processes affecting the groups and clusters properties (see e.g. \citealt{voit05} for a review). A precise calibration of these scaling relations over the full range in mass and redshift is mandatory for high precision cosmology through the groups and clusters mass function. The use of large numbers of objects would indeed lose its interest in the presence of any remaining and not corrected bias in the final relations. At the cluster scale, numerous works have converged towards well defined scaling laws up to relatively high redshifts. At the group scale, such precise calibrations are more difficult to achieve owing the fact that groups present a much wider range of properties at a given mass, i.e. large intrinsic scatters (e.g. \citealt{osmond04,giodini09,rykoff08b,balogh11a}). Moreover, precise mass measurements at this scale are much more difficult to perform regardless of the methodology employed, thus increasing the uncertainties on the scaling relation fits.\\
From the astronomical point of view, these scaling laws are also of great interest as they can be used to put constraints on the underlying physical processes. Observational results can indeed be compared to hydrodynamical simulations of cluster formation to study the relative influence of several mechanisms that modify the ICM properties, such as radiative cooling, supernovae and active galactic nucleus feedback or pre-heating of the gas (see e.g. \citealt{voit05}). On the other hand, a precise characterization of group- and cluster-masses provides the unique way to probe relevant mechanisms such as galaxy harassment, ram pressure stripping or galaxy starvation/strangulation, which drive the evolution of the galaxy properties (stellar mass, size, star formation rate, spectral/morphological type, ...) as a function of their local environment, from field galaxies to the core of massive clusters \citep{smail98,balogh99,tran03,treu03,dressler04,poggianti04,boselli06,poggianti06,balogh07,delucia07,jeltema07,popesso07,huertas09,lubin09,stott09,wilman09,balogh11b,carollo12}. It is, therefore, of main importance to use representative groups and clusters samples covering a large range in mass and redshift to study and constrain the galaxy and ICM properties along with robust and direct mass estimates of the parent halo. This is the main goal of our study: the analysis of a sample of galaxy groups in order to constrain different optical scaling relations, which can be used for cosmological purposes. This study also provide a secure sample of intermediate-mass range objects to investigate the galaxy properties in more details and compare them to field and cluster galaxies.\\
To construct large samples of groups and clusters of galaxies, one can use several methods: spectroscopic identification of galaxy over-densities (e.g. \citealt{miller05,knobel09,cucciati10}), optical detection based on the red sequence galaxies (e.g. \citealt{gladders05,koester07a}), detection of variations in the cosmic microwave radiation due to the SZ effect (e.g. \citealt{carlstrom02,staniszewski09,planck1}), detection of the ICM diffuse X-ray emission (e.g. \citealt{mulchaey98,bohringer00,finoguenov07,vikhlinin09}), observation of weak gravitational lensing distortions of background galaxies (e.g. \citealt{marian06,gavazzi07,massey07,berge08}). Each of these techniques has its own advantages and limitations. For instance, SZ detections are less redshift dependent than X-ray observations that are limited to the high-mass end of the mass function when going to high redshifts. Weak lensing lose its efficiency to low-mass objects and high redshifts ones but is also insensitive to the dynamical state of the target. Optical detections can probe a large range in mass and redshift but suffers from contamination due to projection effects. Spectroscopic redshifts are potentially powerful to construct large samples of groups and clusters but this method requires large quantities of observing time. For one who wants to target low-mass objects up to high redshift, the strong lensing signal produced in the core of some dark matter halo is an interesting alternative. Although such strong lensing events remain rare, their theoretical distribution in terms of angular separation (e.g. \citealt{oguri06}) has been probed over a wide range of halo mass, from galaxy-scale (e.g. \citealt{munoz98,myers03,bolton06,more11}) to cluster-scale objects (e.g. \citealt{luppino99,ebeling01,zaritsky03,gladders03,wen11}). With an automatic search of a large sky area, the observed number of such strong lensing systems will increase, making this detection method interesting (e.g. with the {\it Large Synoptic Survey Telescope}, \citealt{LSST}).\\
At intermediate mass-scales, galaxy groups have been largely investigated with optical (including strong lensing) and X-ray tracers (e.g. \citealt{mulchaey98,helsdon00,zabludoff00,helsdon03,helsdon03b,osmond04,willis05,jeltema06,finoguenov07b,rasmussen07,mamon07,faltenbacher07,gastaldello07,fassnacht08,yang08,giodini09,sun09,cucciati10,leauthaud10,balogh11b,connelly12}). \cite{more12} have presented the most up-to-date sample of objects detected by their strong lensing signal. Its main specificity resides in a selection designed to focus on strong lenses at the galaxy group scale. The study we present here is therefore, in combination with the previous work of \cite{sl2s}, the first analysis of a large sample of strong lensing galaxy groups up to high redshift.\\
This Paper I is organized as follows. In section 2 we present the SARCS sample of lens candidates. We briefly recall our weak lensing methodology in section 3, along with the results of the shear profile fitting. Section 4 is dedicated to the optical analysis of the sample: selection of the bright red galaxies, estimates of richnesses and optical luminosities, luminosity maps and the morphological classification. In section 5, we combine results from the weak lensing and optical analyses to draw a sample of 80 most secure candidates and study the optical scaling relations. We finally draw some conclusions in section 6. A Paper II (Foex et al., in preparation) will focus more on the properties of the galaxy population and correlations with their environment.\\
Throughout this paper, we use a standard $\Lambda$-CDM cosmology defined by $\Omega_\mathrm{M}=0.3, \Omega_\Lambda=0.7$ and a Hubble constant $H_0 = 70\,\mathrm{km/s/Mpc}$.

\section{The SARCS sample}

\subsection{The CFHTLS survey}

The Canada-France-Hawaii Telescope Legacy Survey (CFHTLS\footnote{http://www.cfht.hawaii.edu/Science/CFHLS/}) is a photometric survey made in five bands $u',g',r',i',z'$ close to the bands of the Sloan Digital Sky Survey \citep{fukugita}. Observations were taken with the CFHT prime focus instrument {\sc MegaPrime} covering a field-of-view of 1 deg$^2$ on the sky with a pixel size of 0.186''. The survey includes two components; a Wide component made of four regions of the sky at high galactic latitudes and low extinction, covering in total 170 deg$^2$, and a Deep component, made of four pencil-beam fields of 1 deg$^2$. One of the Deep fields (D1) is located within its Wide counterpart (W1). After masking unusable areas (bright stars and other defects), the CFHTLS survey covers an effective area of 150.4 deg$^2$.\\
The raw images were pre-reduced at CFHT with the elixir pipeline\footnote{http://www.cfht.hawaii.edu/Instruments/Elixir/} and then astrometrically calibrated, photometrically inter-calibrated, resampled, stacked and released by the Terapix group at the Institut dÕAstrophysique de Paris (IAP). We use the CHFTLS T0006 release, in which the Deep fields are offered in two stacks, D-25, which combines the 25\% best-seeing individual pointings, and D-85 using the 85\%. Both the detection of the lens candidates and the weak lensing analysis were done on the D-25 images as they provide a smaller seeing. $i'$-band images that we used for the weak lensing analyses have a seeing $\le0.65''$ for the Deep fields, going up to $0.9''$ for the Wide fields. Typical completeness magnitudes are $m_{i'} = 25\,$mag (Deep) and  $m_{i'} = 24\,$mag (Wide). More details on the T0006 release can be found on the Terapix website\footnote{http://terapix.iap.fr/cplt/T0006-doc.pdf}.

\subsection{The SL2S-ARCS sample}

The Strong Lensing Legacy Survey (SL2S, \citealt{cabanac07}) is an semi-automated search of strong lensing systems on CFHTLS Deep and Wide fields. The SL2S lens sample was compiled using two detection algorithms optimized for different classes of strong lensing systems. The {\sc ringfinder} is an object-oriented  color-based algorithm searching for galaxy-scale lenses around ellipticals. The {\sc ringfinder} produced the SL2S RING sample (Gavazzi et al., in prep.). The {\sc arcfinder} \citep{alard06,more12}, is a generic algorithm aimed at detecting elongated and curved features anywhere in the CFHTLS images, thus more efficient to find group and cluster-scales lenses. The scan of the complete CFHTLS survey resulted in the SL2S-ARCS sample (SARCS) fully described in \cite{more12}.\\
Basically, {\sc arcfinder} search FITS images for elongated and contiguous features of pixels above a given intensity threshold, and tag the most promising features as arc candidates according to their width, length, area and curvature (see Table 1 of \citealt{more12}). On CFHTLS fields, {\sc arcfinder} thresholds were kept low to favor completeness over purity. This led to roughly 1000 candidates/deg$^2$. Then, the candidates were inspected visually, reducing the sample to 413 candidates ($\sim2.75$ candidates/deg$^2$). These potential lenses were then ranked separately by three people, from 1 to 4, 4 being most likely a strong lensing system. The final SARCS sample was extracted from this ranked sample selecting candidates reaching rank 2 or higher with an arc radius $\mathrm{R_{A}}\gtrsim2"$ in order to filter out galaxy-scale systems (the arc radius, defined as the distance between the candidate lensed image and the centre of the respective lens galaxy, is a reasonable proxy for the mass of strong lensing systems). In total, 127 systems were selected, whose general properties are given in Table 2 of \cite{more12}. The redshift distribution of the sample, derived from the photometric redshifts of \citealt{coupon09}, spans a range $z \in [0.2$-$1.2]$ and peaks at z $\sim$ 0.5 (Figure 7 of \citealt{more12}). As seen on Figure 10 of \cite{more12}, the distribution of the image separation of the SARCS systems is located between the galaxy-scale SLACS sample \citep{bolton06} and the massive cluster MACS sample \citep{ebeling01}, thus corresponding mostly in groups and poor clusters of galaxies.\\
We added to the SARCS sample an extra group-scale lens discovered in a different {\sc MegaCam} observation from the CFHTLS fields. This group was part of the previous sample analyzed in \cite{sl2s} (SL2S J09413-1100) and is referenced in the following as SA0. We also removed two candidates from the initial SARCS sample, SA21 and SA56, because of the presence of a large foreground galaxy close enough to the central galaxy to make its color determination not reliable, which is problematic for the optical analysis (richness, luminosity and morphology). In total, we then have a sample of 126 lens candidates. We will use both the weak lensing and the optical analysis of this initial sample to provide a sub-sample of the most secure SARCS group candidates (see Section 5.1).

\section{Individual weak lensing measurements}

\subsection{Methodology}
The weak lensing pipeline we used is fully described in \cite{foex12}. A similar methodology was already employed by \citealt{sl2s} on the first SL2S sample of groups and by \citealt{bardeau07,soucail12} on several galaxy clusters. We outline in the following the main steps of the extraction of weak lensing masses from the CFHT observations.

\subsubsection{Galaxy selection}

First, we detect the objects on the images with {\sc SExtractor}\footnote{http://www.astromatic.net/software/sextractor} \citep{bertin96}. We then use the i-band photometric properties to construct a catalog of stars and a catalog of galaxies. The distinction is made through a combination of the size of the objects with respect to the Point Spread Function (PSF), their position in the magnitude/central-flux diagram (i.e. with respect to the {\it star branch}) and their stellarity given by {\sc SExtractor}. After this first step, we obtain typical number densities of $3\,\mathrm{arcmin^{-2}}$ for stars and $65/25\,\mathrm{arcmin^{-2}}$ for galaxies in the Deep/Wide fields.\\
From the galaxy catalogs, we select the lensing sources as follows. We remove all galaxies within the red sequence (see section 4.1 for its definition) down to $m_{i'}<23$ mag, a limit large enough to reject most of the faint galaxies of the groups without rejecting too many faint field galaxies with similar colors as the group members. Then, we keep only the remaining galaxies with $21<m_{i'}<m_{comp}+0.5$, where $m_{comp}$ is the 50\% completeness limit of the galaxy catalogs in the $i'$-band. The lower limit close to the completness magnitude ensures to keep a control on the redshift distribution of the selected galaxies (required to estimate the lensing strength, see below), while the upper limit $m_{i'}>21$ does a good compromise between removing the foreground galaxies without rejecting too many lensed galaxies. In doing so, we get final densities of roughly $40/15\,\mathrm{arcmin^{-2}}$ galaxies in the Deep/Wide fields.

\subsubsection{Shape measurements}

Next, we estimate the shape of the galaxies using the software {\sc Im2shape}\footnote{http://www.sarahbridle.net/im2shape/} \citep{bridle02} as done in many other studies (\citealt{cypriano04, bardeau05, bardeau07, limousin07a, limousin07b}). Our implementation of {\sc Im2shape} follows exactly the one presented in \cite{foex12}. We use one elliptical Gaussian to model the light distribution of the stars and galaxies to derive their shape parameters. The estimation of the PSF field is made by taking the average shape of the five nearest stars at each galaxy position. The MCMC sampler returns for each galaxy the most likely ellipticity components along with robust statistical errors. With the STEP1 simulations reproducing ground-based observations \citep{heymans06}, this implementation of {\sc Im2shape} was found to present a lensing bias of $\sim-10\%$ \citep{foex12}, a value that is accounted for in this work by increasing the measured shear by 10\%.

\subsubsection{Shear profiles}

\begin{figure}
\center
\includegraphics[width=\hsize, angle=0]{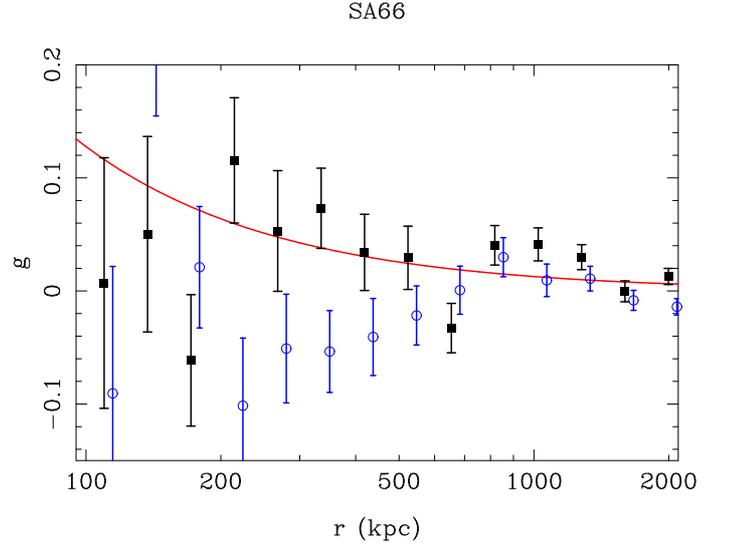}
\caption{Shear profile for the candidate SA66 ($\sigma_{SIS}=644$ km/s, z\_{spec}=0.35). The black filled squares shows the tangential component of the shear, the blue open circles the B-mode of one of the shifted profiles (for clarity). The red curve is the SIS best fit to the tangential shear.}
\label{fig:shear} 
\end{figure}

Once the shape parameters of the galaxies were estimated, we used them to construct shear profiles. Assuming circular symmetry of the lens mass distribution and a random orientation of field galaxies, the average shape of background galaxies in a region of constant potential gives an estimate of the reduced shear $<e>=g$. Thus, the average tangential ellipticity of galaxies in concentric annuli around the lens provides a shear profile $g(r)$ that can be fitted by analytical models to estimate the mass. To reduce the impact of galaxies with a noisy shape estimation, we weighted the ellipticity of each galaxy by the inverse of its variance added in quadrature to the intrinsic shape noise, i.e. the width of the galaxies intrinsic ellipticity distribution ($\sigma_{int}=0.25$, e.g. \citealt{brainerd96}). The lensed galaxies were binned in logarithmic annuli, starting at 50 kpc from the centre (see below for its definition) and with a ratio of 1.25 between the outer and inner limits of the bin. These logarithmic profiles ensure to get a roughly constant signal-to-noise ratio of the shear in each bin along with a good spatial resolution in the central parts. All the profiles were fitted within $r\in[100\mathrm{kpc}-2\mathrm{Mpc}]$. We chose to use a fixed range for all the SARCS groups to avoid over-estimations of the mass by only selecting the region where the signal is significantly positive. We could have used a fixed range of angular radius, but given the large coverage in redshift of the SARCS sample, it would have led to very different regions according to the redshift of the object. As we are mainly studying galaxy groups, the outer limit of 2 Mpc was large enough to probe the shear signal beyond the Virial radius.\\
As described in \cite{foex12}, we constructed several {\it shifted} shear profiles to reduce sampling effects in the central bins where the number of galaxies is small. In practice, we moved the inner part of the first bin by a fraction of its width (e.g. 1/50, so 50 shifted profiles), giving a new estimation of the shear at the corresponding position. Each of these shifted profiles are correlated, but not fitted together, so a classical $\chi^{2}$ minimization still holds. For each of these profiles, we draw 1000 Monte Carlo new profiles (assuming a gaussian distribution for the shear estimates in each bin). This led to a large number of estimates for the best fit parameters of the model (50 times 1000). We used the final distribution to derive the best model, i.e. the mode of the distribution, along with robust statistical errors. An example of the measured shear profile in given in Figure \ref{fig:shear}.

\subsubsection{Normalization of the profiles}

Before we run the fitting procedure of the shear profiles, we have to adjust their normalization. Here we have to deal with two effects. The first one is the contamination of the sources catalogs by un-lensed galaxies. Despite our selection criteria, we expect to have a significant fraction of foreground galaxies in our catalogs, thus implying a dilution of the shear signal. Note that we also have some contamination by group members. However, because of their small number in galaxy groups and thanks to the removal of the red sequence, they do not have a significant effect on the shear measurements. We checked that, in most cases, the number density profile of the lensing sources is roughly flat in the central parts, meaning that the number of remaining group members (compared to field galaxies) in the lens catalogs is negligible. In our case, this possible contamination does not require any correction as done for instance for rich galaxy clusters (e.g. \citealt{hoekstra07,foex12}).\\
The second effect is the determination of the strength of the shear signal to relate the observed distortions into physical quantities. Both are treated simultaneously via the determination of the geometrical factor $D_{ls}/D_{s}$, ratio of the angular diameter distances between the lens and the source $D_{ls}$ to the distance of the source $D_{s}$. We follow here the same methodology used in \cite{sl2s} and \cite{foex12}, which is based on the photometric redshifts catalog from the T0004 release of the CFHTLS-Deep survey. The main advantage of this catalog is that the data were taken with the same instrument in the same photometric system than the present observations. The photometric redshifts we used are the publicly available catalog provided by Roser Pello\footnote{http://www.ast.obs-mip.fr/users/roser/CFHTLS\_T0004/}, redshifts that have been derived with {\sc HyperZ} \citep{bolzonella00} and carefully calibrated with spectroscopic samples \citep{ienna06}. So, following previous works (e.g. \citealt{cypriano04,hoekstra07,sl2s,oguri09,foex12}), we applied directly on this catalog the same color and magnitude criteria used to select the lensed galaxies in order to get a similar redshift distribution. This point requires to neglect the cosmic variance, which is not always valid at the scale of 1 deg$^{2}$ fields of view. We checked indeed that the redshift distribution of the D2 field (part of the COSMOS field, known to be over-dense) gives slightly different results on the average $D_{ls}/D_{s}$ than the D1, D3 and D4 distributions. In the following, we used only the distribution of photometric redshifts from the D1 field.\\
Setting $D_{ls}/D_{s}=0$ for the galaxies with $z_{phot}<z_{lens}$, we then computed the geometrical factor (i.e. the shear strength) averaged over the full redshift distribution. In doing so, we account at the same time for the contamination by foreground galaxies in the source catalogs. We therefore simply translate the dilution of the observed shear signal by unlensed galaxies into the estimate of its strength through the distances ratio. This average geometrical factor can be inverted into an effective redshift $z_{eff}$ such as $\langle D_{ls}/D_{s}\rangle = D_{l,z_{eff}}/D_{z_{eff}}$. These effective redshifts are given in Table 2, and because of the contamination by field galaxies, they are much lower than typical values used elsewhere to derive the strength of the shear signal of low-z clusters, i.e. $z_{s}\sim1$ (e.g. \citealt{okabe08,radovich08}). As verified in \cite{sl2s}, the photometric redshifts from R. Pello give consistent results compared to those obtained with the catalog of \cite{coupon09}.\\
Instead of assuming a typical source redshift, using a geometrical factor averaged over the whole redshift distribution is a better way to convert the observed shear into physical quantities. But it remains an approximation since the reduced shear is not linear in $D_{ls}/D_{s}$ (e.g. \citealt{seitz1997,hoekstra00}). However, we start to fit the shear profiles at 100 kpc from the centre, a distance where the convergence $\kappa$ is subcritical and low enough for group-scale haloes to reduce the influence of this approximation.

\begin{figure}
\center
\includegraphics[width=\hsize, angle=0]{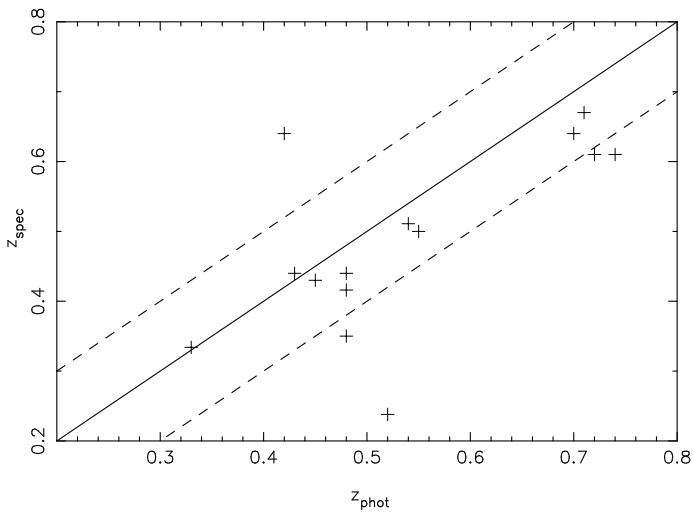}
\caption{Spectroscopic redshifts versus photometric redshifts for the SARCS objects (see Table 2). The solid line shows the equality, the two dashed lines are at $\pm0.1$, i.e. a typical uncertainty on photometric redshifts.}
\label{fig:redshifts} 
\end{figure}

\begin{figure}
\center
\includegraphics[width=\hsize, angle=0]{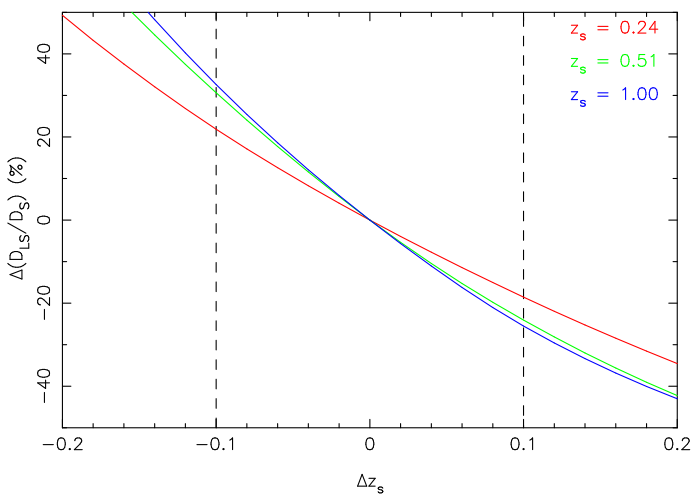}
\caption{Variation of the average geometrical factor $\langle D_{ls}/D_{s}\rangle$ as a function of the shift around the true lens redshift for three groups, SA48 (red curve), SA50 (green curve), and SA123 (blue curve). The two vertical dashed lines mark the typical $\pm0.1$uncertainty of photometric redshifts.}
\label{fig:dls} 
\end{figure}

A last point that needs to be verified is the accuracy of the redshifts we use to estimate $D_{ls}$, along with the impact on the derived lensing masses. Although we have a spectroscopic redshift for some of the SARCS objects (see Table 2), we used in most cases the photometric redshift of the central galaxy derived by \cite{coupon09}. First, we simply compared the spectroscopic and photometric redshifts of the 14 SARCS candidates having both values. As seen on Figure \ref{fig:redshifts}, we obtained an overall good agreement, only two objects presenting a difference larger than 0.2. The first one, SA33, has its central galaxy falling in a masked region. However, the distribution of photometric redshifts (estimated with HyperZ) in the central part of the lens and corrected for the field distribution peaks at $z\sim0.65$, i.e. a value in perfect with the spectroscopic redshift $z_{spec}=0.64$. For the second catastrophic error, SA48, a large and bright foreground galaxy $\sim15$'' away from the lens galaxy might contaminate its photometric redshift estimation. Given the low number of members for this lens, we cannot apply the same procedure as for SA33 with a field-subtracted photometric redshift distribution. In such cases, lens in a masked region, nearby contaminating galaxy or close arc, estimating the redshift of the lens with only the central galaxy can be problematic.\\
The shear signal is fitted by $\gamma=\langle D_{ls}/D_{s}\rangle\times f(M)$, where $f(M)$ is a function of the lens mass $M$. Therefore, an (over-) underestimation of the lens redshift will increase (decrease) $D_{ls}$ and  translate into an (over-) underestimation of its mass. According to our methodology to derive $\langle D_{ls}/D_{s}\rangle$, two effects are adding here when changing the redshift of the lens: the variation of the distance $D_{ls}$, and the value of the contamination level, i.e. the number of galaxies in the D1 catalog matching the selection criteria of lensed galaxies and having $z_{phot}<z_{lens}$. As both effects have a strength that depends on the lens redshift, we tested the procedure described above in three different cases, for a low-redshift group (SA48, $z_{spec}=0.24$), one at the peak of the SARCS sample n(z) (SA50, $z_{spec}=0.51$), and one at high redshift (SA123, $z_{phot}=1.00$). For each group, we derived the average geometrical factor corresponding to different shifts of the lens redshift around its original value, results presented Figure \ref{fig:dls}. As we can see on this plot, for a given variation $\Delta z_{s}$, the change in $\langle D_{ls}/D_{s}\rangle$ is larger for a lens a higher redshift. However, the main variation occurs from low to medium redshifts, as the variations observed for SA50 (z=0.51) and SA123 (z=1.00) are very similar. For a typical $z_{phot}$ uncertainty of $\Delta z_{s}=\pm0.1$, we would get an error of $20\%-30\%$ on the geometrical factor, thus a mass over-/under-estimation of the same order. Assuming that the $z_{phot}$ for most of the SARCS lenses is accurate up to this 0.1 precision, the precise knowledge of $z_{s}$ introduces an error lower than the statistical one (quoted in Table 2) due to the dispersion of the galaxy intrinsic ellipticities. As detailed below, we used for the weak lensing analysis the singular isothermal sphere mass model, which has a shear function $\gamma\propto(D_{ls}/D_{s})\times\sigma_{v}^{2}$. In that case, a variation less than $30\%$ in $D_{ls}/D_{s}$ induces a change less than $20\%$ in the velocity dispersion $\sigma_{v}$. Therefore, the intrinsic ellipticity dispersion remains the principal limitation in our weak lensing analysis. However, on specific cases with larger $\Delta z_{s}$, we obtain a significant bias. For instance, SA48 has $z_{phot}=0.52$, and $z_{spec}=0.24$, so a geometrical factor underestimated by $\sim40\%$. Using the photometric value to fit the shear profile would have led in that case to a velocity dispersion $\sigma_{v}^{phot}=\sqrt{1/0.6}\times\sigma_{v}^{spec}$, i.e. $\sim30\%$ larger than the value derived with the spectroscopic redshift.

\subsection{Choice of the centre}
To construct a shear profile, we need to specify its centre, which should correspond to the centre of the mass distribution. With enough constraints, its position can be considered as a free parameter of the mass model and fitted during the weak lensing analysis. It is also possible to use 2D pixelized mass reconstructions and associate the highest projected density peak to the mass centre. However, as we are dealing here mainly with group-scale dark matter haloes, the strength of the shear signal is not strong enough to do that. A few attempts to fit both the mass profile and the centre (using the {\sc Lenstool} code, \citealt{kneib96,jullo07}) on some cases have shown that no robust constraint can be obtained from the weak lensing signal we have. We also tried to reconstruct 2D mass maps, but in most cases they were too noisy to derive the location of the mass centre. Moreover, \cite{dietrich11} have shown from a set of ground-based simulated data that, on 2D mass distributions derived from weak lensing, shifts between the real centre and the reconstructed one have a median value of 1', therefore limiting seriously the use of this method to derive a reliable position of the mass centre.\\
By its definition, the SARCS sample has the advantage of giving an idea where the highest density regions are located. The SARCS lens candidates are indeed detected by a strong lensing feature with an arc radius corresponding to a strong lensing event by a group-scale halo (the selection threshold of $RA\gtrsim2"$ limits the presence in the sample of strong lensing events by a galaxy-scale halo enhanced by a group-scale halo, see e.g. \citealt{sl2s}). Thus we can assume that the strong lensing system is a good indicator of the position of the actual mass centre as a high enough mass density is required, i.e. $\kappa\gtrsim1$. However, we expect that this assumption might be wrong in some cases where the lens is not a regular and isolated dark matter halo but rather presents a complex morphology. For instance, \cite{limousin10b} showed that the SARCS lens SA66 (SL2S J08544-0121 in their paper) is a clear bimodal object spectroscopically confirmed by \cite{munoz13}, for which the mass centre is not associated to the centre of the strong lensing system. As shown in section 4.2, there is a non-negligible number ($\sim15\%$) of such complex and multimodal systems in the SARCS sample.\\
The other option usually taken when no strong lensing arcs unambiguously identify the centre of the halo consists to assume that the brightest galaxy in the dark matter halo lies in the centre of its mass distribution. The models of formation of large cD galaxies, e.g. infall and merging of galaxies on the central one \citep{ostriker75,hausman78} or accretion of the intra cluster gas due to the cooling flow in the centre of the gravitational potential well \citep{cowie77}, predict that such object are found indeed in the centre of their host halo. This hypothesis is also supported by observational results where the cD galaxy is found in the kinematical centre, e.g. \cite{quintana82}. However, the brightest galaxy is not always a cD at rest in the gravitational potential well, but rather a large elliptical galaxy that is not necessarily located at the centre of the mass distribution. For instance, \citealt{jeltema07} have studied a sample of 7 X-ray loud galaxy groups at intermediate redshifts ($0.2<z<0.6$) and for two of them, the brightest galaxy present an offset $\sim100h_{70}^{-1}\,\mathrm{kpc}$ with the X-ray emission peak. Note that they also found two groups where none dominant elliptical galaxy is present, but rather several ones with comparable luminosities (see also the more recent work by \citealt{george12} of the COSMOS X-ray selected groups). In such cases, the assumption of tracing the mass centre by the position of the most luminous galaxy might be wrong as well. This problem, sometimes referenced as the central galaxy paradigm (e.g. \citealt{skibba10} for a review), is clearly a limitation to build an accurate shear profile from the position of the brightest member in a halo. In our case, it seems wiser to use the strong lensing system to trace the position of the mass centre, and we will use it for all objects in the sample, regardless if the lens galaxy is the brightest one in the halo. In some cases, the strong lensing system is indeed associated to a satellite galaxy; we discuss the impact of using the strong lensing system as centre instead of the brightest galaxy for such groups in the Appendix.\\
Finally, it is worth to mention that the mass underestimation due to small miscenterings can be simply limited by not using the central parts of the shear profiles in the fitting process \citep{mandelbaum10}. As we start to fit the profiles at 100 kpc, we therefore avoid such possible underestimations in most cases.
 
 \begin{figure}[!t]
\center
\includegraphics[width=\hsize, angle=0]{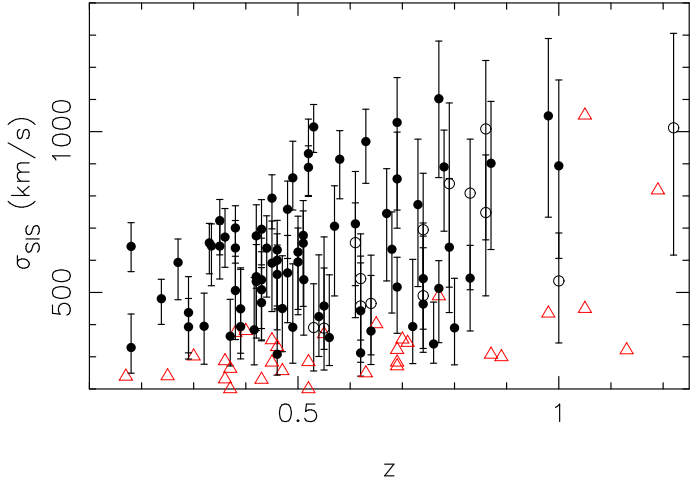}
\caption{Velocity dispersion derived from the fit of the shear profile using the SIS model. Red open triangles show the upper limit on $\sigma_{v}$ for those objects not having a $1\sigma$ weak lensing detection. Open circles are candidates with a weak lensing detection less than $3\sigma$, filled circles are those having a reliable detection (above $3\sigma$).}
\label{fig:sigma2z} 
\end{figure}
 
\subsection{Results}
Observed shear profiles are fitted using the singular isothermal sphere model (SIS hereafter), which describes the mass density of a relaxed massive sphere characterized by a constant velocity dispersion $\sigma_{v}$. The lensing functions write $\gamma(r)=\kappa(r)=\mathrm{R_{E}}/2r$, where $\gamma(r)$ is the shear, $\kappa(r)$ is the dimensionless projected mass density, and $r$ the projected distance to the lens centre. The Einstein radius scales as $\mathrm{R_{E}\propto(D_{ls}/D_{s})\sigma_{v}^{2}}$. The (one-dimensional) velocity dispersion $\sigma_{v}$ is used as the free parameter to fit the SIS model, and not the Einstein radius as it requires an estimate of the source redshift. As mentioned previously, the signal to noise ratios that we measure are in most cases too low to get reliable information on the properties of the mass distribution. So we did not try to fit, for instance, the widely used NFW model \citep{navarro97,navarro04}, thus avoiding poorly constrained results. We emphasize that the SIS model is only employed to derive a raw mass estimate and not probe the shape of the mass profiles. As shown by \cite{oguri06} (see also \citealt{more12}), the range of image separations probed by the SARCS sample corresponds to lensing events produced by a mix of SIS (low-mass end) and NFW (high-mass end) haloes. Most of the SARCS candidates are supposed to be galaxy groups and present an arc radius compatible with a SIS lens. We therefore introduce in our weak lensing analysis a bias due to the SIS modeling only for the few massive galaxy clusters in the sample. But even in these cases, given the large statistical noise we have on the lensing measurements, the SIS approximation does not result in significant variations on the total mass. An alternative would have been to stack the objects and increase the quality of the signal, as done for instance in \cite{mandelbaum06,johnston07,leauthaud10,okabe10a,oguri12}. However we are more interested here in the weak lensing detection of each SARCS objects rather than a precise analysis of the mass distribution at the group scale, as it can also be achieved with a combination of the weak lensing signal at large scale with a strong lensing modeling of the central mass distribution and a dynamical analysis of the group members (e.g. \citealt{verdugo11}). The stacking analysis of this sample and the characterization of the mass profile will be presented in Paper II.\\
On the other hand, the SIS model gives results that can be easily compared to other methods to estimate the mass such as dynamical analysis (e.g. \citealt{munoz13}). Moreover, in the case of the SARCS sample, it is straightforward to compare the weak lensing Einstein radius to the observed arc radius $\mathrm{R_{A}}$, which is equivalent to the actual Einstein radius for axisymmetric lenses. Values are given in Table 2, where the weak lensing $\sigma_{SIS}$ are converted in $\mathrm{R_{E}(z_{s},\sigma_{SIS})}$ with a source redshift $z_{s}$ derived from the redshift distribution of \cite{more11} with the CFHTLS T0006 release i-band limiting magnitude $m_{lim}=24.48$. For some objects, the difference between $\mathrm{R_{A}}$ and $\mathrm{R_{E}}$ is significant, suggesting either an inaccurate weak lensing estimation or a complicated mass distribution of the lens that affects the $\mathrm{R_{A}}-\mathrm{R_{E}}$ relation (strong lensing associated to a satellite galaxy, large ellipticity/asymmetry of the lens, substructures, ...). A deeper study of some of these cases using strong lensing modeling will be presented in Verdugo et al., in prep.\\
From this systematic analysis of the whole SARCS sample, we obtained constraints at the $1\sigma$ level on the SIS velocity dispersion for 89 candidates ($\sim71\%$ of the sample). In the rest of the paper, we will call these objects 'weak lensing detections'. For the remaining objects, the fit of the shear profile only returns an upper limit on $\sigma_{v}$, and we will not use them in the rest of the analysis (objects labelled further as 'non-detected'). Using a $3\sigma$ level cut to select the weak lensing detections leads to a sample of 75 objects. However, the goal here is not to select the most secure lenses but rather remove the most likely false detections. We checked for instance that some of the objects having a detection level between $1\sigma$ and $3\sigma$ present an obvious optical luminosity over-density in the luminosity map (Section 4.2) along with a clear strong lensing system. That is why we chose here a rather loose selection criterion to be combined in Section 5.1 with the optical selection criterion. To calibrate the scaling relations in Section 5.3 we will however use only objects with a $3\sigma$ weak lensing detection level.\\
The distribution of the SARCS candidates in the $z-\sigma_{v}$ plane is shown Figure \ref{fig:sigma2z}. The average velocity dispersion of the 89 weak lensing detections is $\langle \sigma_{v}\rangle = 618 \pm 197\,\mathrm{km\,s^{-1}}$, corresponding to a rich group, or a poor cluster, depending where the boundary between the two regimes is drawn. From Figure \ref{fig:sigma2z}, we see that the distribution $\sigma_{v}(z)$ is fairly homogeneous over the redshift range. We detect however more massive objects above $z=0.5$. From a weak lensing analysis it is expected as, for a source at a given redshift, the lensing strength decreases for lenses at higher redshifts. Therefore it is normal to observe a larger fraction of more massive systems at higher redshifts. We also see on Figure \ref{fig:sigma2z} that these high redshift objects present larger error bars on the velocity dispersion, owing the lower density of available background galaxies to measure the shear signal. It is interesting to notice that we do not observe a strong trend between the non-detected objects and their redshift. This lack of correlation suggests that the intrinsic quality of the ground-based optical images (seeing, pixel size) that we used is the main limitation to detect low-mass objects, rather than the noise induced by lower densities of background galaxies.\\
Finally, it is worth to mention that at the low-mass end of the sample, the shear signal is weak enough to be close to the noise level. Hence, our weak lensing results for such objects have to be taken with caution. We emphasize here that we are more interested in the detection of the objects rather than accurate mass measurements. At the group-scale, we probably lose some 'real' objects and have as well some false detections or galaxy-scale systems among our 89 objects. The cross-checking with the optical properties (Section 5.1) shows, for instance, that some of the weak lensing detections have no optical counter-part, corresponding to such cases where our lensing procedure fits noise.\\
Looking in more details to the properties of these non-detections, it appears first of all that they mainly correspond to SARCS candidates with a small arc radius. We have indeed 29 objects with $\mathrm{R_{A}}<4''$, i.e. $\sim78\%$ of the non-detections. The whole SARC sample contains 88 objects with $\mathrm{R_{A}}<4''$ ($\sim70\%$ of the sample), a slightly lower value simply reflecting the fact that less massive objects (i.e. with smaller arc radius) are more difficult to detect via weak lensing. We only have two candidates not detected in weak lensing with $\mathrm{R_{A}}>5''$, one of them having a large arc radius, SA104 with $\mathrm{R_{A}}=11.7''$ ($\mathrm{z_{phot}}=0.15$). This object, ranked 2 in \cite{more12}, shows a single very large elliptical galaxy without any obvious companion around. Given the arc radius of $\sim$12'', this object should have much more members as it would correspond to a poor cluster. We then can safely consider this object as a false detection in the SARCS sample. The other object is SA41 ($\mathrm{R_{A}}=6.1''$, $\mathrm{z_{phot}}=0.52$). This object presents a bimodal light distribution. A wrong choice of the centre to compute the shear profile could be the reason of the weak lensing non detection. However, using a different centre does not improve the constraints (see Appendix).\\
Among the 37 non-detections, we have 25 candidates ranked less than 3, i.e. $\sim68\%$. This value is slightly higher than the fraction obtained for the entire sample, which contains $\sim57\%$ of such low ranked candidates. It suggests that the threshold $rank\ge2$ used to build the initial SARCS sample is not strong enough to prevent keeping false detections. In Section 5.1, we discuss the properties of the most secure candidates in terms of the initial selection parameters (rank and arc radius).\\
Beside systems falsely identified as strong lenses, one can think of several hypotheses to explain why objects with a small arc radius are not detected with the weak lensing method. First, these strong lensing features are most likely produced by less massive systems (or even galaxy-scale haloes), which do not produce a weak lensing signal strong enough to be detected with ground-based data. Second, a star field that is too sparse to properly sample the PSF across the field-of-view can reduce the quality of the galaxy shape estimation, hence introducing more noise in the signal. On the other hand, a field with too many large diffracted stars has a smaller effective area to measure the shear signal, which can bias the weak lensing analysis. Third, the intrinsic mass distribution of the lens might as well be a strong factor of noise in the measured shear signal. We use indeed the simplest weak lensing analysis assuming circular symmetry. In the case of highly elliptical mass distributions, such an approximation can result in a significant underestimation of the shear. It can also generate apparent small arc radius, not representative of the total mass of the halo, thus explaining why we have weak lensing detections for some candidates with small $\mathrm{R_{A}}$. For multimodal systems, the question of the centre and the fit to a single halo might as well bias the mass determination. In the Appendix, we present some results for such complex systems, along with some cases where the strong lensing system is not associated to the brightest galaxy in the halo but to a satellite galaxy. We emphasize here all these possible reasons just to remind that a weak lensing detection is sufficient but not necessary to conclude that we are observing a massive halo.


\subsection{Comparison with previous mass measurements}

Several SARCS groups have been already analyzed using different data sets and methodology:
\begin{itemize}
\item weak lensing: with the CFHTLS T0004 release, \cite{sl2s} derived weak lensing constraints on $\sigma_{SIS}$ only for 5 groups of their sample of 12 objects. With these 5 objects, we find an average ratio $\langle\sigma_{WL}/\sigma_{SIS}\rangle=0.93\pm0.10$ (where $\sigma_{SIS}$ are the velocity dispersions derived in this work and $\sigma_{WL}$ those from \citealt{sl2s}). Note that there is only one group in the sample of \cite{sl2s} for which we did not get a weak lensing detection, SA122 ($\mathrm{z_{phot}}=0.69$, $\mathrm{R_{A}}=2.8$, rank=3.0), and for which \cite{sl2s} only obtained an upper limit on $\sigma_{SIS}$.

\item strong lensing: 8 objects were analyzed by \cite{sl2s} using the CFHTLS ground-based images, and 4 new groups have been studied with HST data (Verdugo et al., in prep.). To compare the results of \cite{sl2s} to ours, we have converted their Einstein radii into $\sigma_{SL}$ assuming the source redshift given in Table 2. With these 8+4 groups, we obtain an average ratio $\langle\sigma_{SL}/\sigma_{SIS}\rangle=0.92\pm0.25$.

\item dynamical analysis: 7 objects have been studied by \cite{munoz13} using VLT/FORS2 spectra. Here we obtain an average ratio of $\langle\sigma_{dyn.}/\sigma_{SIS}\rangle=0.61\pm0.25$.
\end{itemize}

Despite our weak lensing methodology follows closely that of \cite{sl2s} (same procedure to select the lensed galaxies and estimate their shape parameters, same normalization of the shear profiles), we managed to measure $\sigma_{SIS}$ for 11 of their sample of 12 groups while they obtained constraints only for 5 of them (we attribute this increased number of detection to our statistical analysis of the shear signal). On the other hand, the results we have are very similar with a ratio of $\sim0.9$ and a small scatter of $10\%$. Our velocity dispersions are also compatible on average with the values derived from the strong lensing analysis of \cite{sl2s} and Verdugo et al. (in prep.), with again a ratio of $\sim0.9$, but with a larger scatter of $\sim25\%$. This seems to indicate that these groups do not present a very large ellipticity with a major axis aligned along the line-of-sight, or, equivalently, a high concentration that could artificially enhance the measured mass when extrapolating the strong lensing constraints to the larger scales probed with weak lensing signal.\\
The comparison of our measurement with the dynamical velocity dispersions derived by \cite{munoz13} is more puzzling. We obtain indeed a ratio of $\sim0.6$ with a scatter of $\sim25\%$. Among the 7 groups, 6 have a dynamical velocity dispersion smaller than the weak lensing one. Only one group, SA72, has $\sigma_{dyn}>\sigma_{WL}$, with compatible values at the $1\sigma$ level given the large lower limit on $\sigma_{dyn}$. \cite{munoz13} argue that one reason of such a discrepancy could arise from the choice of the SIS model to characterize the actual mass distribution of the groups. With numerical simulations, they also show that mass estimates derived from the velocity dispersion of galaxies in a halo can be underestimated up to 20\%. This is, however, not enough to explain the discrepancies observed here. Another possibility to account for the differences between the lensing and dynamical results would be the presence of massive structures along the line of sight. As the weak lensing signal is produced by all the projected matter between the lensed galaxies and the observer, groups and clusters of galaxies or even large scale structures can affect the shear signal and induce and overestimation of the mass. Using the Millennium Simulation, \cite{hoekstra11} found that randomly positioned massive structures do not statistically bias the weak lensing mass estimate of a galaxy cluster but rather increase its uncertainty, with values comparable with those due to the intrinsic dispersion of the galaxies ellipticity. Most of the overestimated masses they obtained have an excess less than 20\%, but going up to a factor $\sim2$ for some objects. Such projection effects could therefore explain the larger weak lensing masses we have. We can also invoke a poor lensing signal-to-noise ratio from which the weak lensing analysis can return biased masses. As the objects analyzed by \cite{munoz13} are mostly low-mass groups with $\sigma_{dyn}<500\,\mathrm{km/s}$, they are indeed not producing a strong shear signal, possibly leading to wrong estimates. However, we observe a similar discrepancy for all objects, which suggests that this systematic difference in the velocity dispersions is due to the methodologies employed, rather than the groups properties or poor constraints. In fact, the estimation of galaxy groups and clusters' velocity dispersion is know to be biased low by several effects (see e.g. \citealt{biviano06} and reference therein) such as the inclusion of interlopers (i.e. infalling galaxies along filaments), the rejection of high-velocity galaxy members, presence of substructures or the so-called 'velocity-bias' (i.e. different velocity dispersions between galaxies and the dark matter). We are currently increasing the number of groups analyzed via the dynamical methodology, and we will explore in more details the discrepancy between the lensing and dynamical estimate of the groups velocity dispersion (Motta et al., in preparation).

\section{Optical properties}

Although the weak lensing results suggest that some of the SARCS candidates are galaxy-scale lenses or false detections, we did the optical analysis for all objects in the sample. We therefore intend to make a cross-correlation of the two analyses to derive a sample of the most bona fide SARCS groups candidates used to constrain the optical scaling relations. 

\subsection{Richness and optical luminosity}
\begin{figure}
\center
\includegraphics[width=\hsize, angle=0]{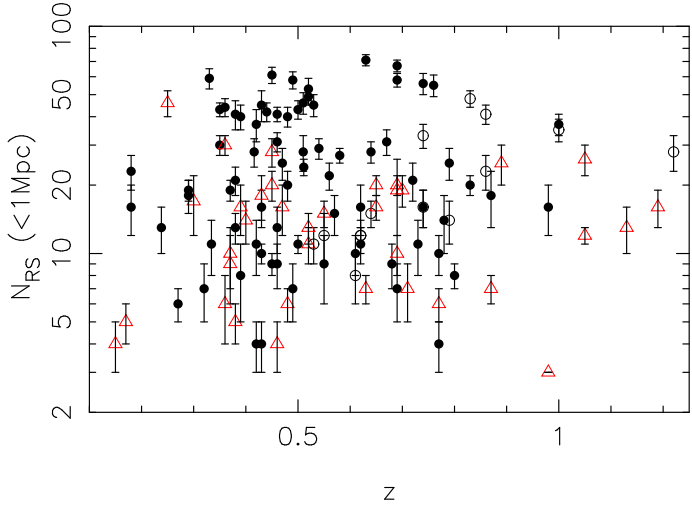}
\caption{Richnesses estimated within 1 Mpc using the bright red galaxies of the 126 SARCS candidates. Open red triangles are objects without a $1\sigma$ weak lensing detection, open circles have a detection between 1 and 3$\sigma$, filled circles are those with a detection above 3$\sigma$.}
\label{fig:n2z} 
\end{figure}

\begin{figure}
\center
\includegraphics[width=\hsize, angle=0]{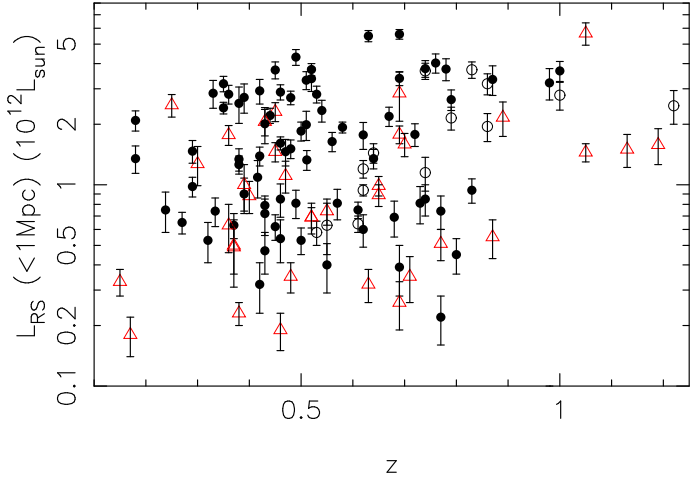}
\caption{Same as Figure \ref{fig:n2z} for the optical i'-band luminosities estimated within 1 Mpc using the bright red galaxies of the 126 SARCS candidates.}
\label{fig:L2z} 
\end{figure}

We derived the optical properties of the SARCS lens candidates from the bright galaxies that belong to the red sequence. Because most of the SARCS objects are groups with a small number of galaxies, we did not attempt to fit this red sequence as it is usually done when dealing with rich galaxy clusters. We used the same criteria for all the candidates and defined the red sequence as the region in the color-magnitude diagram where galaxies have a $r'-i'$ color close to that of the 'lens' galaxy, i.e. the one at the centre of the strong lensing system. In the cases where this lens galaxy is not the brightest one but rather a satellite galaxy, we used the color of the former to define the red sequence. To account for the expected slope of the red sequence (e.g. \citealt{stott09}), we chose asymmetric limits and select galaxies with $(r'-i')_{lens}-0.2<r'-i'<(r'-i')_{lens}+0.15$. As said previously, the galaxy at the centre of the strong lensing system is not necessarily the brightest one. So its color $(r'-i')_{lens}$ can be slightly different from that of the brightest member which is usually taken as reference. However it gives a robust estimator of the group members color since, by definition, it belongs to the group. Because the colors are derived from magnitudes estimated in a fixed aperture of 3'', this $(r'-i')_{lens}$ color tends to be underestimated for systems presenting an arc radius $\mathrm{R_{A}}\leq3''$ where the lens galaxy is close to the strong lensing feature (a blue arc in most cases). For these objects, we used the average color of the surrounding bright galaxies that are most likely part of the system.\\
We restrict the red sequence to an absolute magnitude $M_{i'}=-21$. In doing so, we roughly probe a constant fraction of the luminosity function, which allows direct comparison from group to group regardless of their redshift. Focusing on the brightest galaxies also avoids the fall out of the completeness magnitude of the CFHTLS observations for groups at high redshifts.\\
In \cite{sl2s}, the group members were visually selected and no background correction was applied. Here, as we adopt an automated approach for all objects, we have to account as well for the contamination by field galaxies. We determined in a reference field the density of galaxies falling in the definition of the red sequence of each group. To keep it simple, we used the 1 deg$^{2}$ image of the field where the group was found (for systems in the Wide part of the CFHTLS survey, we only used the central pointing as reference). To get a rough estimate of the fluctuations due to local over/under-densities in the distribution of field galaxies, we computed the background density in 1000 circular patches randomly positioned in the reference field. The size of these patches is chosen to match the area where group members are counted, i.e. within a projected radius of 0.5 and 1 Mpc from the central galaxy. We have chosen to compute the optical properties within two different radius in order to test its influence on the calibration of the scaling relations.\\
To reduce the impact of overestimating the local density of field galaxies, which can lead to a negative number of galaxies for poor groups (as well for systems falsely identified as group), we divided the area where galaxies are counted in concentric annuli. The background subtraction is done in each of these annuli and we finally take the sum only of the positive counts excesses. The richnesses we derived, i.e. the number of galaxies within our selection criteria, are the average of these sums over the 1000 values of the background densities. The scatter around this average gives a rough estimate of the corresponding statistical error. We did the same to compute the optical luminosities accounting for both the k-correction and the passive evolution of an elliptical galaxy (values derived from the synthetic SED model of \citealt{bruzual03}). For consistency, we applied this method to all the SARCS candidates, wether they are false detections, poor groups or poor clusters for which a usual background subtraction works fine. The distribution of richnesses and optical luminosities for the SARCS objects are given in Figure \ref{fig:n2z} and Figure  \ref{fig:L2z}. Within an aperture of 1 Mpc around the centre of the strong lensing system and cutting the luminosity function at $M_{i'}=-21$, the sample covers richnesses up to 70 galaxies and luminosities up to $\sim6\times10^{12}\mathrm{L_{\odot}}$. Both distributions are roughly homogeneous in redshift, and are dominated by group-scale objects with $N\sim5-20$ and $L\sim0.5-1.5\times10^{12}\mathrm{L_{\odot}}$.\\
We chose to derive the optical properties of the SARCS candidates from the galaxies within their red sequence. These galaxies are indeed easier to detect and select (stronger contrast with the population of field galaxies), and most of studies about the optical scaling relations make use of this specific population of early-type galaxies. However we would like to emphasize here that such a selection can introduce some systematics in the analysis. It has been observed indeed that at higher redshifts, groups and clusters contain a larger fraction of blue star-forming late-type galaxies \citep{butcher84,ellingson01,lubin02}, with on the other hand a smaller fraction of red passive early-type galaxies \citep{smail98,kodama04,delucia07}. This evolution of the red sequence, where the blue spirals evolve into red elliptical galaxies with a transient state of green galaxies \citep{balogh11b}, naturally introduces a bias in our galaxy selection as a function redshift as we used a fixed broad red sequence for all objects. Another possible source of systematics in the estimation of richnesses and optical luminosities is the actual fraction of the galaxy population that inhabit the red sequence.  For instance, \cite{zabludoff98} have studied a sample of 12 nearby poor galaxy groups and found significant variations (up to a factor 2) in the fraction of early-type galaxies. Similar variations in the galaxy population from group to group have been obtained by \cite{jeltema07} at intermediate redshifts. Therefore, we are not probing the same fraction of group members for each object in our sample. However, given the relatively large size of the SARCS sample and its broad redshift range, we expect to average such effects (intrinsic variations and redshift evolution), and so we did not attempt to correct them, or to include the green and blue galaxies in the analysis.

\subsection{Morphological classification}
\begin{figure}
\center
\includegraphics[width=\hsize]{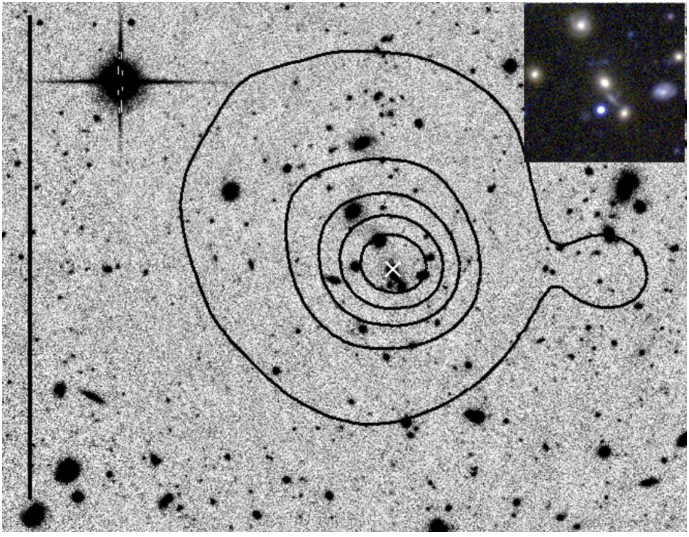}
\caption{Luminosity density contours (in black) for SA15 equal to $10^{6}$, $4\times10^{6}$, $7\times10^{6}$, $10^{7}$, and $1.3\times10^{7}\,\mathrm{L_{\odot}.kpc^{-2}}$. The white cross marks the galaxy at the centre of the strong lensing system. The black vertical line on the left is 1 Mpc long. SA15 is at $z=0.44$. The stamp in the top-right corner shows a 30''$\times$30'' color image of the system.}
\label{fig:SA15} 
\end{figure}

\begin{figure}
\center
\includegraphics[width=\hsize]{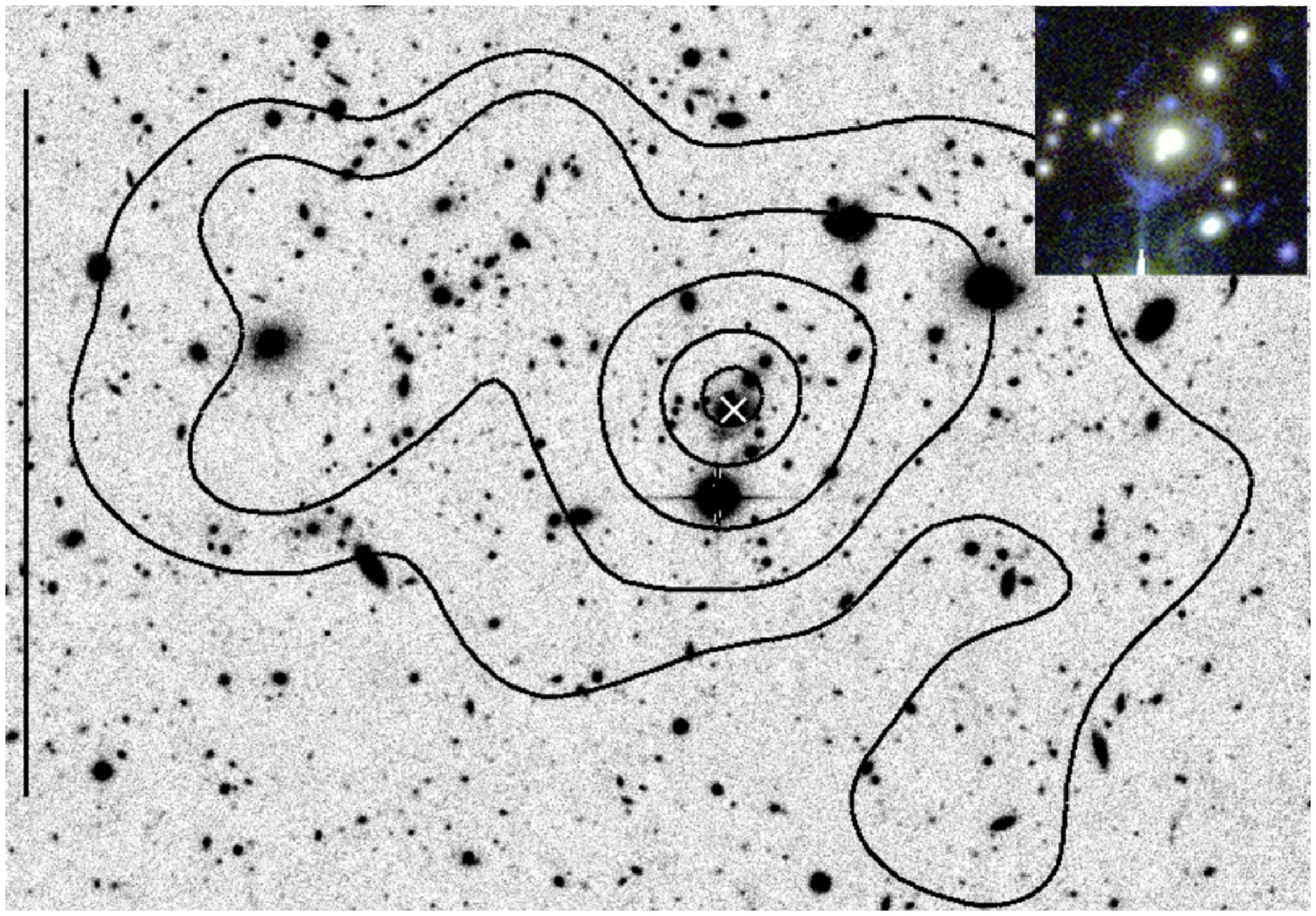}
\caption{Luminosity density contours (in black) for SA2 equal to $2\times10^{6}$, $4\times10^{6}$, $7\times10^{6}$, $1.5\times10^{7}$, and $2\times10^{7}\,\mathrm{L_{\odot}.kpc^{-2}}$. The white cross marks the galaxy at the center of the strong lensing system. The black vertical line on the left is 1 Mpc long. SA2 is at $z=0.48$. The stamp in the top-right corner shows a 30''$\times$30'' color image of the system.}
\label{fig:SA2} 
\end{figure}

\begin{figure}
\center
\includegraphics[width=\hsize]{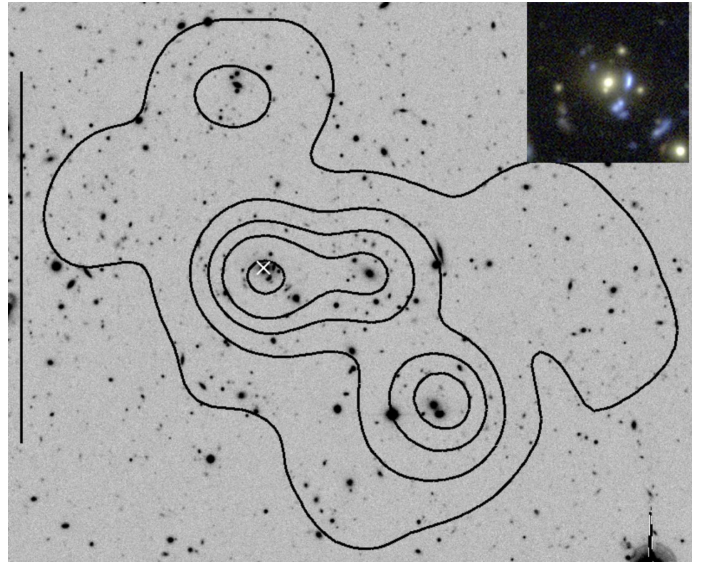}
\caption{Luminosity density contours (in black) for SA90 equal to $1.5\times10^{6}$, $4\times10^{6}$, $7\times10^{6}$, $10^{7}$, and $1.5\times10^{7}\,\mathrm{L_{\odot}.kpc^{-2}}$. The white cross marks the galaxy at the centre of the strong lensing system. The black vertical line on the left is 1 Mpc long. SA90 is at $z=0.53$. The stamp in the top-right corner shows a 30''$\times$30'' color image of the system.}
\label{fig:SA90} 
\end{figure}

From the catalogs of galaxies falling in the red sequence (to get more galaxies and less statistical noise when drawing the luminosity contours, we pushed the limiting magnitude to $M_{i'}=-20$ instead of -21), we computed luminosity maps following \cite{sl2s}. The $15'\times15'$ field-of-view around the lens is divided in cells of $20\times20$ pixels. From the centre of each of these cells (the pixels of the luminosity maps), we looked for the 5 nearest galaxies belonging to the red sequence, a value small enough to avoid oversampling. The luminosity density of the corresponding pixel is simply the sum of the luminosity of these 5 galaxies divided by the circular surface covered by the furthest one. The maps of the luminosity density are then smoothed by a gaussian kernel with a FWHM of 7 pixels ($\sim 25''$). We checked that the shape of the resulting maps are weakly dependent of the pixel size or the smoothing width.\\
As stated previously, we did not clean the catalogs of the red galaxy members from field galaxies. So the maps suffer from the background contamination, but we can assume it to be roughly homogenous across the field, thus not leading to strong shape distortions of the group/cluster itself. However, this can be wrong for groups with low numbers of galaxies in the red sequence. Despite our adaptive smoothing, the classification between a regular or elongated object can be indeed affected by statistical noise due to local variations of the density of field galaxies. On the other hand, because of the large width of the red sequence we used, we expect to pick up over-densities of galaxies that are not necessarily linked to the initial target (i.e. not a multimodal object). In fact, this can be used as a tool to trace the cosmic web and reveal large scale structures around galaxy groups (Cabanac et al., in prep.).\\
Once the maps were built, we visually inspected them to assess the luminous morphology of the SARCS objects according to the shape of the luminosity contours, which levels were adapted for each object. However, we checked that the value chosen for the innermost luminosity contour does not influence the occurrence of high luminosity peaks, i.e. the multimodal groups definition. We sorted the groups according to their morphology in 4 classes:
\begin{itemize}
\item false detection or galaxy-scale strong lensing feature (i.e. no clear over-density in the map) $\rightarrow$ 30 objects
\item regular (i.e. roughly circular isophotes around the strong lensing system)  $\rightarrow$ 39 objects
\item elongated (i.e. elliptical isophotes with a roughly constant position angle form inner to outer parts)  $\rightarrow$ 40 objects
\item multimodal (i.e. 2 or more peaks in the central part of the map) $\rightarrow$ 17 objects
\end{itemize}
Figure \ref{fig:SA15}, \ref{fig:SA2}, and \ref{fig:SA90} present the luminosity map for a regular group (SA15), an elongated group (SA2), and a bimodal group (SA90). Multimodal class refers here only to two or more peaks in the luminosity map found within a 0.5 Mpc radius of the strong lensing system. Extending this limit to a larger radius would increase the number of objects in this class, e.g. 23 members if we look up to 1 Mpc from the centre. However, in these cases we are most likely observing two distinct objects (or an ongoing merging event) rather than a single halo as, given the mass range of these groups, the Virial radius is expected to be $\lesssim1$ Mpc (see e.g. \citealt{munoz13}). We look for a trend between the morphological class and the redshift or mass of the objects in section 5.1, after defining the final sample of the best candidates.\\
From this qualitative morphological classification, it appears that the main part of the SARCS candidates are groups or poor clusters with irregular shapes, either elongated or more complex, i.e. $(40+17)/96\sim60\%$ of the optically detected lenses. This suggests that groups of galaxies are mostly in a young dynamical state. In the context of the large scale structure formation and evolution, this is somehow expected as groups are continuously forming and merging into more massive clusters (e.g. \citealt{evrard90,bekki99}). In paper II will be presented a more quantitative analysis of the groups morphology, along with the study of correlations to the groups environment.

\section{Combining the weak lensing and optical analyses}

\subsection{Selection of the most secure candidates}

As mentioned in Section 2.2, the thresholds applied to the {\sc arcfinder} algorithm were chosen to favor completeness over purity. Despite the visual ranking performed by three different persons, the final SARCS sample still contains galaxy-scale lenses and even some false detections. Both the weak lensing and the optical analyses have indeed shown that some objects do not reach our criteria to be selected as a group-scale lens.\\
From the weak lensing analysis, we end up with a reduced sample of 89 objects with a weak lensing detection. The rejected objects are either false detections, not massive enough haloes (very poor groups or galaxy-scale lenses), or objects with a too noisy shear signal to derive a secure SIS velocity dispersion (sparse data, morphology too complex for a simple spherical mass model, ...). As said in Section 3.3, they present a larger fraction of low-ranked and with smaller arc radius $\mathrm{R_{A}}$ objects than in the total sample.\\
From the visual inspection of the color images and the luminosity maps, the initial SARCS sample get reduced to 96 objects among which 39 present regular isophotes, 40 elongated ones, and 17 have a multimodal configuration. Here, we rejected all the candidates for which we do not observe a clear over-density of light associated to the strong lensing system, i.e. objects where the lensing feature is in a poor environment without evidence of a population of galaxies with similar colors. As for the weak lensing selection, this optical selection mainly rejects SARCS candidates with a small arc radius, i.e. probably galaxy-scale objects or very poor group lenses. Only four rejections are associated to arc radius $\mathrm{R_{A}}>3''$ and most likely correspond to false detections, e.g. edge-on spiral galaxies.\\
While the optical selection removes 30 objects, the weak lensing selection rejects 37 candidates, so a similar number of possible lenses. Interestingly, the two methods have 21 rejected candidates in common, which are most certainly not group-scale lenses and can be securely removed from the final sample. On the other hand, we have 16 candidates not detected in weak lensing but flagged as probable groups from their luminosity maps. Among them, only 4 objects have regular isophotes, which suggest that we do not measure a good enough shear signal because of the complex morphology of the mass distribution, i.e. multimodal or highly elliptical. We also have 9 objects for which we managed to put constraints on $\sigma_{SIS}$ but which do not have an obvious optical counterpart (our richness estimator gives for all of them a null or negative value). We visually inspected these 'dark lenses', and for four of them we found a significant galaxy concentration less than 5' away from the supposed strong lensing system. In these cases, the shear signal that we measure is most likely due to a close (in projection) massive structure not associated to the SARCS candidate. In two other cases, the PSF map derived from the field of stars show a non-smooth pattern that might generate a false shear signal. For the three remaining objects, we could not find any obvious explanation for the measured shear signal given that the optical images clearly show the absence of a galaxy concentrations around the SARCS candidate.\\ 
Finally, the combination of our selection criteria leads to a sample of 80 lenses ranging from group to cluster-scale haloes. Their weak lensing and optical properties are given in Table 2. In terms of the morphological distribution of this sample of most secure lenses, we have 34 objects with regular isophotes ($\sim42\%$), 33 with elongated/elliptical ones ($\sim42\%$), and 13 multimodal groups ($\sim16\%$) with a second luminosity peak closer than 0.5 Mpc from the strong lensing system. The different ratio are roughly similar to those obtained for the 96 candidates having a clear optical detection, and our final sample still contains a large fraction of objects with an irregular light distribution, i.e. $\sim57\%$. The average velocity dispersions in each morphological class are all compatible within their $1\sigma$ statistical scatter as we obtain $592\pm175\,\mathrm{km\,s^{-1}}$ for the regular groups, $589\pm201\,\mathrm{km\,s^{-1}}$ for the elongated ones, and $716\pm147\,\mathrm{km\,s^{-1}}$ for the multimodal, i.e. a value slightly larger. We also looked for any redshift trend, but the three classes have a very similar average redshift.\\
The initial sample has $\sim70\%$ of objects with $\mathrm{R_{A}}<4''$ (observed $\mathrm{R_{A}}$, not derived from $\sigma_{SIS}$) and $\sim57\%$ objects ranked less than 3, i.e. the threshold used in \cite{more12} to define the most promising candidates. In our final sample we obtain fractions of $\sim63\%$ and $\sim49\%$: our optical and lensing criteria result in a sample with a larger fraction of promising candidates (based only on the visual inspection of the strong lensing features) and with larger arc radius. If we assume that the best group- and cluster-scale lens candidates can be a priori defined as those having both a rank $\ge3$ and $\mathrm{R_{A}}\ge4''$, then our final sample contains $18/20$ of the best candidates in the initial SARCS sample, which suggest that this two criteria are pretty robust to select such real lenses at the group-scale.\\
To reduce the impact of unreliable measurements, we will keep only the objects with a $3\sigma$ weak lensing detection to fit the scaling relation. This subsample of the most secure candidates according to our combined weak lensing and optical analysis contains 67 objects. In doing so, we lose some objects at high redshifts, without improving the dispersion in richness or optical luminosity (see Figures \ref{fig:sigma2z}, \ref{fig:n2z}, and \ref{fig:L2z}). As we have 14 objects (13 with an optical confirmation, among which 2 have a spectroscopic confirmation and a strong lensing model) with a weak lensing detection level between 1 and 3$\sigma$, we lose $\sim16\%$ of the 80 lenses subsample defined here. Therefore, this sample with a larger statistic, especially at high redshift, will be used in other works to study the population of galaxy groups, e.g. Verdugo et al. (in prep.).


\subsection{Scaling relations at the group scale}

\begin{figure*}
\center
\includegraphics[width=18cm, angle=0]{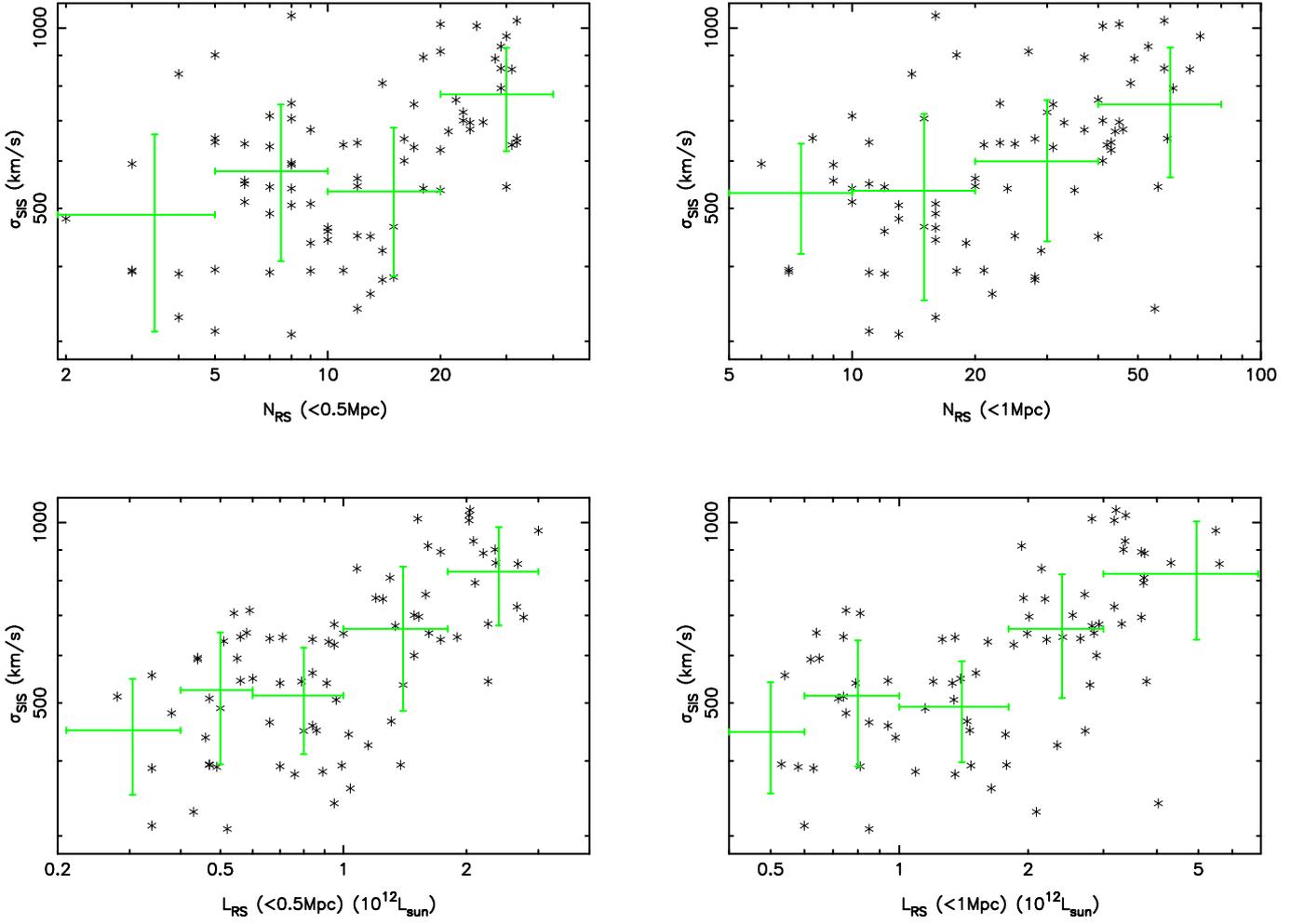}
\caption{Velocity dispersion derived from weak lensing as a function of optical richness (first row) and luminosity (second row) estimated with the bright red galaxies in 2 apertures, 0.5 Mpc (left column) and 1 Mpc (right column). We used only the sample of the 80 candidates defined in Section 5.1 (error bars on each individual measurement are omitted for clarity, see Table 2). Green points with error bars highlight the increase of $\sigma_{v}$ with richness and luminosity after binning the SARCS lenses according to their observed richness or luminosity.}
\label{fig:nL} 
\end{figure*}

We used the sample of the 80 most secure candidates as defined previously to look for correlations between the mass derived from weak lensing and the optical properties. Such scaling relations, characterized by power laws, have been observed at different mass scales and redshifts (e.g. \citealt{lin03,lin04a,popesso05,brough06,becker07,johnston07,popesso07,reyes08,mandelbaum08,rozo09,andreon10,foex12}). Usually, scaling relations are investigated using spherical NFW mass at a given density contrast, e.g. $M_{200}$, as they are related to the total Virial mass. Because the SARCS sample is mainly made of galaxy groups, we kept our weak lensing analysis to its simplest version with only estimates of the SIS velocity dispersion. As the SIS model is already a significant approximation of the actual mass distribution, we did not use SIS masses in a given aperture as it would increase the scatter of the correlations, but simply used the SIS velocity dispersions. Moreover, due to the lack of information on the actual mass profile of the lenses, we do not have estimates of their Virial radius, although \cite{munoz13} give a raw estimation for some of the groups. Therefore, we used richnesses and luminosities derived in fixed physical apertures (0.5 and 1 Mpc) regardless of the mass and the redshift of the objects.\\
Our results are presented Figure \ref{fig:nL}. In all cases, we observe a quite large dispersion in $\sigma_{SIS}$ with scatters ranging from 15\% to 35\%, without any obvious trend as a function of richness or luminosity. However, when the objects are binned according to their richness or luminosity, we detect the expected correlations, the more massive objects being more luminous and with more galaxies populating their red sequence.\\
It is interesting to note in both right panels of Figure \ref{fig:nL} the presence of a clear outlier in the bottom-right corner. This object is embedded in a large scale structure extending over several Mpc. It is located close to the node of this filamentary structure, so when counting galaxies up to 1 Mpc, we face a contamination by the surrounding clumps of galaxies. We also observe three outliers for the $N-\sigma_{v}$ relations with an apparently over-estimated velocity dispersion given the richness ($\sigma_{v}>800\,\mathrm{km\,s^{-1}}$, $N_{0.5Mpc}<10$). However, these objects do not appear as outliers in the $L-\sigma_{v}$ relations. In two cases, the presence of two bright galaxies of similar magnitude in the centre can explain this behavior (for the third one, another bright galaxy with similar colors falls into the 0.5 Mpc region from the lens). However, owing the large size of the red sequence, we cannot securely discriminate between group members and field galaxies from which projections effects could explain the observed large luminosity given the richness of these three objects (projection effects could also be responsible here for over-estimated lensing masses).\\
A more accurate calibration of these relations will be presented in Paper II, where groups will be stacked and outliers removed, e.g. objects in large scale structures, or with a very disturbed light morphology.

\subsection{From poor groups to rich clusters}

\begin{figure}
\center
\includegraphics[width=\hsize, angle=0]{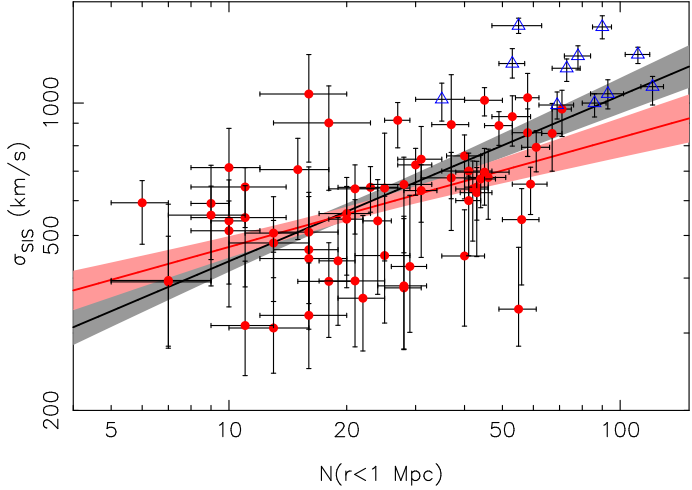}
\caption{Richness-velocity dispersion scaling relation. The black solid line shows the best BCES orthogonal fit of the $N_{1Mpc}-\sigma_{SIS}$ relation using the SARCS most secure candidates (red points) and adding the EXCPRES galaxy clusters of \cite{foex12} (blue open triangles). The grey shaded area gives the statistical 1$\sigma$ uncertainty given by the best fit parameters (statistical dispersion $\sigma_{stat}$). The red line (and corresponding shaded area) is the best fit using only the SARCS most secure candidates.}
\label{fig:Nsigma} 
\end{figure}

To get a larger range in mass, we combined this sample of the best SARCS candidates with the sample of rich and massive galaxy clusters presented in \cite{foex12}. These 11 clusters are part of the EXCPRES sample ({\it Evolution of X-ray galaxy Cluster Properties in a REpresentative Sample}, Arnaud et al. in preparation), which was designed as the REXCESS sample \citep{bohringer07} in order to study the evolution of the X-ray properties of clusters. The full EXCPRES sample contains 20 clusters in the redshift range $0.4<z<0.6$, observed with XMM-{\it Newton}. Only clusters with an X-ray luminosity $L_{X} > 5\times10^{44}\,\mathrm{erg/s}$ in the [0.5-2.0] keV band within the detection radius were selected for an optical follow-up at the CFHT. The X-ray properties and the results of the weak lensing analysis of this sub-sample of 11 clusters are presented in \cite{foex12}.\\
To be consistent with the work on the SARCS sample, we re-analyzed the optical images ({\sc Megacam} data) of the EXCPRES clusters and applied exactly the same procedure used here to get the richnesses and luminosities. In total, we have 67+11 objects to adjust four scaling relations, $\sigma_{SIS}-N_{0.5\,\mathrm{Mpc}}$, $\sigma_{SIS}-N_{1\,\mathrm{Mpc}}$, $\sigma_{SIS}-L_{0.5\,\mathrm{Mpc}}$, and $\sigma_{SIS}-L_{1\,\mathrm{Mpc}}$ (we used single measurements to fit the correlations, not binned values as in other works such as \citealt{reyes08,leauthaud10}). Some of the SARCS candidates are located close to the edge of the MegaCam field-of-view, at a projected distance smaller than the size of the aperture used to derive the optical properties. We removed these objects from the fit of the scaling laws, i.e. one object for the correlation at 0.5 Mpc and four for those at 1 Mpc (objects notified by a $^{*}$ in Table 2).\\
In order to get more quantitative results, we fitted the correlations using the bootstrapping orthogonal BCES estimator ({\it Bivariate Correlated Errors and intrinsic Scatter}, \citealt{akritas96}) as done in some previous works, e.g. \cite{morandi07,pratt09,foex12}. The main advantage of this approach, compared to simple linear regression, is that it accounts for the intrinsic dispersion of the objects around the best fit. This dispersion needs to be included and evaluated in the fit. It gives indeed a crude idea of the impact of some physical processes that cause a departure from the theoretical predictions. For instance, the large intrinsic dispersion observed in the mass-X-ray luminosity scaling relation is a good tracer of the physics that take place in clusters of galaxies such as radiative cooling, pre-heating or feedback from supernovae (e.g. \citealt{voit05} for a review).\\
To optimize the fit of the scaling relation and reduce the correlation between its logarithmic slope and normalization, we normalized both variables by a pivot representative of the sample: 10 and 20 for richnesses in 0.5 Mpc and 1 Mpc, $10^{12}\,L_{\odot}$ and $2\times10^{12}\,L_{\odot}$ for luminosities. The velocity dispersions are normalized by 600 $\mathrm{km\,s^{-1}}$. The results of the BCES estimator are given in Table 1 and Figure \ref{fig:Nsigma} shows the best BCES fit for the $\sigma_{SIS}-N_{1\,\mathrm{Mpc}}$ scaling law. As in \citet{lin04a,foex12}, we also give the value of the Spearman correlation coefficient $\rho$ which shows the degree of monotony of a given correlation.

\begin{figure}
\center
\includegraphics[width=\hsize, angle=0]{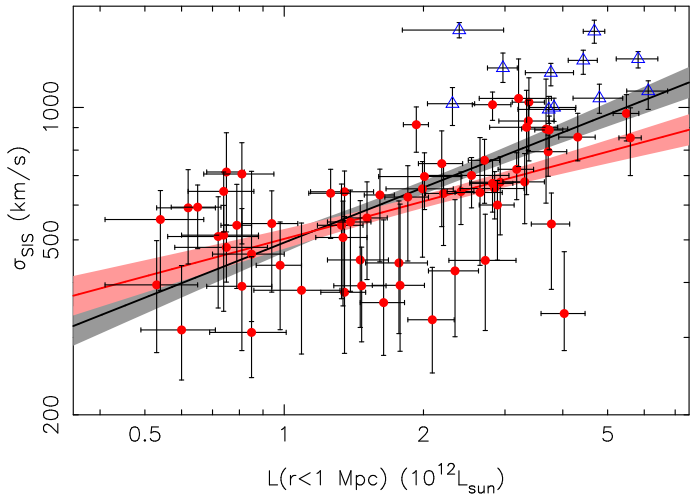}
\caption{Same as Figure \ref{fig:Nsigma} but for the Luminosity-velocity dispersion scaling relation.}
\label{fig:Lsigma} 
\end{figure}
 
\begin{table*}
\label{table:scaling}
\centering 
\begin{threeparttable}
\caption{Summary of the fitting results for the scaling relations $(\sigma/\sigma_{0})=A\times(Obs./P_{Obs.})^{\alpha}$.}
\begin{tabular}{l c c c c c c c c}
\hline\hline\noalign{\smallskip}
scaling law & $\alpha$ & $A$ & $\sigma_{raw}$ & $\sigma_{stat}$ & $\sigma_{int}$ & $\rho$ & R & $P_{Obs.}$\\
\noalign{\smallskip}\hline\noalign{\smallskip}
$\sigma_{SIS}-N_{0.5\,\mathrm{Mpc}}$ & $0.36\pm0.05$ & $0.96\pm0.04$ & $0.13\pm0.02$ & $0.06\pm0.01$ & $0.11\pm0.02$ & $0.65$ & 24\% & 10\\
$\sigma_{SIS}-N_{0.5\,\mathrm{Mpc}}$ (SARCS only) & $0.24\pm0.05$ & $0.93\pm0.03$ & $0.11\pm0.01$ & $0.08\pm0.01$ & $0.03\pm0.18$ & $0.68$ & 24\% & 10\\
\noalign{\smallskip}
$\sigma_{SIS}-L_{0.5\,\mathrm{Mpc}}$ & $0.45\pm0.05$ & $1.04\pm0.03$ & $0.13\pm0.01$ & $0.07\pm0.01$ & $0.11\pm0.02$ & $0.70$ & 24\% & $10^{12}\,L_{\odot}$\\
$\sigma_{SIS}-L_{0.5\,\mathrm{Mpc}}$ (SARCS only) & $0.32\pm0.05$ & $0.98\pm0.03$ & $0.10\pm0.01$ & $0.08\pm0.01$ & $0.01\pm0.40$ & $0.76$ & 24\% & $10^{12}\,L_{\odot}$\\
\noalign{\smallskip}\hline\noalign{\smallskip}
$\sigma_{SIS}-N_{1\,\mathrm{Mpc}}$ & $0.38\pm0.05$ & $0.94\pm0.04$ & $0.12\pm0.02$ & $0.06\pm0.01$ & $0.10\pm0.02$ & $0.65$ & 24\% & 20\\
$\sigma_{SIS}-N_{1\,\mathrm{Mpc}}$ (SARCS only) & $0.25\pm0.06$ & $0.93\pm0.04$ & $0.11\pm0.01$ & $0.08\pm0.01$ & $0.06\pm0.12$ & $0.68$ & 25\% & 20\\
\noalign{\smallskip}
$\sigma_{SIS}-L_{1\,\mathrm{Mpc}}$ & $0.42\pm0.06$ & $1.10\pm0.04$ & $0.13\pm0.01$ & $0.07\pm0.01$ & $0.11\pm0.02$ & $0.67$ & 25\% & $2\times10^{12}\,L_{\odot}$\\
$\sigma_{SIS}-L_{1\,\mathrm{Mpc}}$ (SARCS only) & $0.28\pm0.06$ & $1.02\pm0.03$ & $0.11\pm0.01$ & $0.08\pm0.01$ & $0.01\pm0.23$  & $0.73$ & 24\% & $2\times10^{12}\,L_{\odot}$\\
\noalign{\smallskip}\hline
\end{tabular}
    \begin{tablenotes}
      \small
      \item Columns are (1) scaling relation. (2) Best fit logarithmic slope. (3) Best fit normalization. (4-5-6) total, statistic, and intrinsic logarithmic dispersions. (7) Spearman coefficient. (8) Relative error on $\sigma_{v}$ using the scaling relation as mass proxy. (9) Pivot chosen to normalize the observable (velocity dispersions are normalized by $\sigma_{0}=600\,\mathrm{km\,s^{-1}}$).
    \end{tablenotes}
  \end{threeparttable}
\end{table*}

First of all, the correlations appear to be roughly the same working at 0.5 or 1 Mpc from the centre. This basically means that the choice of the aperture in which richnesses and luminosities are estimated is not a dominant parameter. This has already been observed in similar studies (e.g. \citealt{andreon10,foex12}) with consistent results using either a fixed aperture or scaling it with mass. Thanks to the large number of objects, we obtain small statistical dispersions, the intrinsic scatter around the best fit being the main source of noise (see Figure \ref{fig:Nsigma}) and \ref{fig:Lsigma}, with $\sigma_{int}\sim0.11$ (dex). The correlations are pretty strong with Spearman coefficients of $\rho\sim0.6-0.7$: both the richness and the optical luminosity of the bright red galaxies in a halo are good tracers of mass. We can assess the quality of our best fits as mass proxies simply by converting richnesses and luminosities into $\sigma_{v,proxy}$ and look at that average relative error over the sample $R=\langle |(\sigma_{v,WL}-\sigma_{v,proxy})/\sigma_{v,WL}| \rangle$ (see Table 1). According to this criterion, our scaling relations are efficient to recover velocity dispersions with an accuracy of better than $\sim25\%$.\\
While the hierarchical model of structure formation predicts a number of sub-haloes proportional to the mass of the parent halo, i.e. $N\propto M$ (e.g. \citealt{delucia04,gao04}), including baryons in numerical simulations results in a decrease of the number of galaxies for halo of higher masses, i.e. $N\propto M^{\alpha}$ with $\alpha<1$ (see e.g. \citealt{white01,berlind03}). Several mechanisms can be responsible of this reduced slope, such as a decrease of gas cooling and star formation efficiency \citep{springel03,borgani04,kravtsov04} or an increased merger/destruction rate of galaxies in higher mass objects \citep{white01,lin03}. Our results for the $\sigma_{v}-N$ relations are in good agreement with theoretical predictions from numerical simulations with a slope of 0.33 (e.g. \citealt{evrard08,lau10}), as well as similar work \citep{andreon10}. However, we do not observe any significant evidence of the influence of physical processes that might affect the galaxy population. For the $\sigma_{SIS}-L$ relations, we obtain slightly larger slopes than the $\sigma_{SIS}-N$ correlation with values not consistent with a constant mass-to-light ratio. This result suggests that the physical mechanisms inside a dark matter halo are more efficient to affect the luminosity rather than the number of red sequence galaxies, including for instance ram pressure stripping or galaxy harassment (see e.g. \citealt{treu03,poggianti04,boselli06}). The presence of a higher fraction of galaxies with evolved stellar populations in higher mass objects has also been invoked to explain the increase of the mass-to-light ratio with mass (e.g. \citealt{bahcall02}). However, this assumption has been ruled out by \cite{popesso07} and their study of the Fundamental Plane of the cluster ellipticals, showing that they have a constant mass-to-light ratio that cannot explain the variations in the total mass-to-light ratio of clusters. We intend to use our spectroscopic surveys of the SARCS groups to investigate this hypothesis in more details and down to lower-mass objects in Paper II.\\
We also fit these scaling relations using only the sample of galaxy groups. We obtained slopes slightly lower but consistent within $2\sigma$ uncertainties with the groups+clusters combined fits (see Table 1; also see Figure \ref{fig:Nsigma}) and \ref{fig:Lsigma}). This result is not surprising as we use bootstrapping to derive the best fit parameters, which gives more weight to the SARCS groups as they are more numerous than the EXCPRES clusters. With the relations calibrated with the SARCS objects only, we derived the associated R value. The $\sigma_{SIS}-L_{1\,\mathrm{Mpc}}$ scaling gives $R=24\%$ (22\% when applied on the SARCS sample only, 37\% for the EXCPRES clusters only), and the $\sigma_{SIS}-N_{1\,\mathrm{Mpc}}$ relation gives $R=25\%$ (23\% and 34\%). These relations give similar accuracies despite lower slopes. This result is mainly due to the different size of the two samples, the SARCS one dominating the statistic. We find larger R values for the EXCPRES sample only, suggesting a break in the scalings between the low- and high-mass ends of the combined sample. However, as we do not obtain statistically significant differences in the slopes of the scaling relations between groups and groups+clusters (discrepancies lower than $3\sigma$), we cannot reliably discriminate between a single population of similar objects over the whole mass range from the presence of two distinct subpopulations.

\subsection{Comparison with previous works}

Although this work is the first one based on such a large sample of strong lensing groups, we are not the first to investigate the group properties through some scaling relations. As stated previously, groups have been already extensively studied with different baryonic tracers, from X-ray or optical observations. Despite some difficulties to make proper comparisons (e.g. different way to define the optical richness, measure the X-ray luminosity, ...), we review here some of these works.\\
From the X-ray point of view, \cite{jeltema06,jeltema07} presented the analysis of a sample of 9 X-ray emitting galaxy groups at intermediate redshifts \citep{mulchaey06}. In particular they found that these intermediate-redshifts groups present similar properties as local groups, thus excluding strong non-standard redshift evolution. The X-ray properties of these objects (temperature, luminosity, entropy) follow the scaling relations obtained with galaxy clusters and/or nearby galaxy groups (see also \citealt{mulchaey98}). In Section 5.1, we showed that we obtained large scatters in the velocity dispersion at a given richness or optical luminosity, a result also observed by \cite{jeltema06} for a given temperature or X-ray luminosity of the gas.\\
More recently, a large sample of ~200 X-ray galaxy groups up to a redshift of 1 has been drawn from the COSMOS field \citep{scoville07} and studied with a weak lensing analysis to put constraints on the mass-X-ray luminosity scaling relation by \cite{leauthaud10}. They found a very good agreement with other studies, i.e. that this correlation can be characterized by a single power-law over a very large range in mass (see also \citealt{foex12}). Despite some hints of a possible break of similarity at the cluster scale, we have shown in the previous section that a single power law gives also an acceptable fit of the scaling with richness and luminosity from poor groups to rich clusters. They claimed to observe a little evidence for a non-standard redshift evolution of the relation, but without a strong statistical significance. Their work therefore also confirm that galaxy groups are following the same scalings as galaxy clusters rather than presenting special properties. \cite{giodini09} studied some galaxy groups and poor clusters observed in the COSMOS field. Their analysis of the stellar mass fraction associated to the galaxy members showed again no evidence of a non-standard redshift evolution as their scaling relation agrees with that obtained for local clusters (see also the work by \citealt{connelly12}). Their results on galaxy groups are also in good agreement with those for galaxy clusters only, supporting again the idea that galaxy groups are not a particular population of halo but just a scaled-down version of clusters, as expected in the simple model of hierarchical structure formation and evolution.\\
As stated in Section 3.2, using the same COSMOS data, \cite{george12} confirmed the findings of \cite{jeltema07} about the BGG/BCG in groups and clusters, i.e. an early-type galaxy, but not necessarily dominant, and with observed shifts up to 100 kpc with respect to the X-ray peak (see also \citealt{hoekstra12} and their sample of 50 massive galaxy clusters). In our study of the SARCS sample, we observed the same behavior in some specific cases such as the multimodal groups where there is not a single early-type galaxy dominating the light distribution, and/or where the strong lensing system is not associated to the brightest member. A more quantitative analysis of this problematic will be presented in Paper II.\\
Numerous works have explored galaxy groups based on optical observations, both from the side of the properties of individual galaxies and as a global population correlated to the host dark matter halo. In this paper, we focused on the second point of view through some optical scaling relations. Our results obtained in Section 5.3 on the calibration of the $\sigma_{v}-N$ and $\sigma_{v}-L$ scaling relations do not give a strong evidence of departure from the purely gravitationally driven model of structure formation, although the mass-to-light ratio we observed is not constant across the range in mass. As mentioned previously, galaxy-galaxy and galaxy-halo interactions are taking place in groups and clusters, and they can modify the global properties of the galaxies populating the halo (see e.g. \citealt{lin04a}). Several studies have been looking for observational evidences of these mechanisms associated to the baryonic content of dark matter haloes. For instance, the maxBCG sample of optically selected groups and clusters observed in the SDSS survey \citep{koester07a,koester07b} has been widely analyzed to derive the optical $M-N$ and $M-L$ scaling laws. \cite{reyes08} have used $\sim13000$ objects in the maxBCG catalog, cutting towards low richnesses at $N_{200}\ge10$ (according to their definition of the red sequence), and covering a range in redshift from 0.1 to 0.3. They binned the objects in several ways, according either to richness, luminosity or luminosity of the BCG. They derived the corresponding stacked weak lensing mass and fitted the optical scaling laws. They obtained consistent behaviors over the entire mass range, with a small evidence of non-gravitational processes, as the slopes of their relations are larger than 1 at the $2\sigma$ level. Similar results were obtained by \cite{johnston07} despite a slightly different sample, i.e. including objects at lower richnesses (see also \citealt{mandelbaum08,rozo09}). However, as noticed in \cite{andreon10}, the sample used by \cite{johnston07} suffers from the Malmquist bias, resulting in smaller slopes. We expect here to have the same issue as of our weak lensing analysis uses the low-mass end of the SARCS sample, where the shear signal gets too low and noisy. In Section 3.3 we showed indeed that 76\% of the weak lensing non-detections are associated to strong lensing systems with an observed arc radius smaller than 4''. Therefore, we most likely miss a significant fraction of the low-mass objects in the sample, the direct consequence on our calibration being a decrease in the fitted slopes.\\
Despite a large number of works with different methodologies, the precise characterization of the optical scaling relations (slopes and scatters) remains an open problem. There is a trend in the different studies with evidences of the role of the baryonic physics through slopes larger than 1, e.g. the works on the maxBCG catalog, or \cite{muzzin07} with $M\propto N^{1.4\pm0.2}$ from 15 clusters, \cite{lin04a} with $N\propto M^{0.82\pm0.04}$ and $L\propto M^{0.72\pm0.04}$ from 93 groups and clusters, \cite{bardeau07} with $M\propto L^{1.8\pm0.24}$ with 10 galaxy clusters (see also \citealt{marinoni02,popesso05,parker05,popesso07}). However some authors still obtain slopes consistent with 1 and so no evidence of the processes affecting the galaxy properties (see e.g. \citealt{andreon10,foex12}). With the present work, we did not rule out the simplest model of structure formation, although our mass-to-light ratio seems to indicate some differences between the population of the most massive haloes and that of small galaxy groups. A more extensive analysis of the galaxy properties in the SARCS groups will be presented in Paper II. We will make use of this sample to investigate where/when are galaxies the most affected, i.e. interaction with the parent halo during the infall, galaxy-galaxy interactions, environment effects, ... (see e.g. \cite{cibinel12} and reference therein).

\section{Conclusions}
In this first paper we presented the weak lensing analysis of the SARCS sample of lens candidates, a work that follows the previous one made by \cite{sl2s} on a first sample of group candidates. These potential groups and clusters of galaxies were detected on the CFHTLS survey by the presence of a possible strong lensing feature. Taking advantage of the high quality of the CFHT images, i.e. deep observations and good seeing, we studied each object one by one, which is done for the first time at the group scale (previous weak lensing studies of galaxy groups were stacking objects). We were able to measure a shear signal for a large part of the sample as we obtained a detection for 89 candidates, i.e. with a SIS velocity dispersion larger than 0 at least at the $1\sigma$ level. As most of the SARCS objects are galaxy groups up to high redshifts ($z=1.2$) with a faint and noisy shear signal, we only focused on estimates of the total mass via the SIS mass distribution instead of trying to assess the radial or 2D mass distribution. The SIS velocity dispersion we obtained for the SARCS sample is dominated by galaxy-group objects with an average value of $\sigma_{v}\sim600\,\mathrm{km\,s^{-1}}$. We also found some galaxy clusters in the sample with velocity dispersions up to $1000\,\mathrm{km\,s^{-1}}$. We did not find strong evidence of correlation between the measured SIS velocity dispersion and redshift, indicating that the SARCS sample is fairly homogenous up to $z\sim0.8$.\\
We also performed the optical analysis of SARCS objects. Using the galaxies belonging to the red sequence down to an absolute magnitude of $M_{i'}=-21$, we estimated for each objects their optical richness and luminosity in different fixed apertures, i.e. not scaled with mass or redshift. We obtained typical values of $\mathrm{N(R<1Mpc)}\sim5-20$ (up to $\sim70$) and $\mathrm{L(R<1Mpc)}\sim0.5-1.5\times10^{12}L_{\odot}$ (up to $\sim6\times10^{12}L_{\odot}$). We also use the catalogs of red galaxies to construct 2D luminosity map and explore the morphology of the SARCS candidates. Our classification resulted in two main conclusions:\\
(i) A significant number of the confirmed groups and poor clusters present complex morphologies such as very disturbed luminosity contours or several luminosity peaks in the central parts. This suggest that groups of galaxies are mostly dynamically young structures.\\
(ii) as for the weak lensing analysis, some of the objects were not clearly detected, then possibly corresponding to galaxy-scale lenses or false detections. The combination of the optical and weak lensing results led us to a final sample of 80 galaxy groups, removing some clear false detections as SA104, thus improving the purity of the sample. Compared to the initial SARCS sample, we obtain a similar morphological distribution. On the other hand, we obtain a larger fraction of high-ranked objects, and with larger arc radius. The selection criteria used in \cite{more12} were indeed selected to favor completeness over purity, which is most likely much higher in our final sample.\\
Finally, we quickly explored the relation between mass and the main optical properties. Despite significant scatters up to 35\% in $\sigma_{v}$ at a given richness or luminosity, we found good correlations as more massive systems are richer and more luminous. We combined the SARCS sample with a sample of rich clusters of galaxies and obtained consistent results over the entire range in mass with obvious scaling relations between the SIS velocity dispersion and the global properties of the galaxies population. With this work, we confirm the possibility to use the optical scaling relations as reasonable mass proxies to analyze large samples of groups and clusters of galaxies and derive cosmological constraints via their mass function. However, our results have to be considered with caution as we performed a lensing analysis on single objects, which can lead to biased results for the low-mass objects, and in most cases to large uncertainties in the mass measurement, which are the current limitation of the statistical significance of the results presented here. In Paper II (Foex et al., in prep.), we will stacks the objects in order to get a more robust weak lensing signal, and thus put tighter constraints on these scaling relations. In doing so, we hope to significantly reduce the systematic uncertainties until the point where the lack of a good understanding of the SARCS selection function will have to be accounted for. We are attempting to assess this problem of the SARCS selection function by conducting a lens search of the complete CFHTLS imaging via a citizen science project (More et al., in prep).\\
This Paper I is only one step in the study of the SARCS sample. Ongoing and complementary observations will provide new results to be compared to this preliminary weak lensing and optical analyses. In particular, multi objects spectroscopy will increase the number of groups presented in \cite{munoz13} for which the mass inferred by a dynamical methodology can be used to test the reliability of the weak lensing results presented here. With a combination of these dynamical results with our weak lensing study and some strong lensing modeling (Verdugo et al., in prep, see also \citealt{verdugo11}), we intend to investigate in more details the mass profile from the central parts of the group up to large radius, thus testing some predictions from numerical simulations. We will also explore more closely the properties of the galaxies inside the groups such as the evolution of the red sequence with redshift, the size of galactic dark matter haloes inside groups with a galaxy-galaxy weak lensing analysis or the halo occupation distribution and the central galaxy paradigm (Foex et al., in prep., Paper II). Thanks to the large field of view of {\sc Megacam} and the large area covered by the CFHTLS survey, a search for large scale structures linked to the SARCS objects is being explored (Cabanac et al., in prep.). We also intend to correlate all these observational results with large N-body dark matter numerical simulation to put constraints on the formation and the evolution of galaxy groups and their link with the large-scale structures of the Universe.

\begin{acknowledgements}
G.F. acknowledges support from FONDECYT through grant 3120160.\\
V.M. acknowledges support from FONDECYT through grant 112074.\\
M.L. acknowledges the Centre National de la Recherche Scientifique (CNRS) for its support. The Dark Cosmology Centre is funded by the Danish National Research Foundation.\\
T.V. acknowledges support from CONACYT grant 165365 through the program ÒEstancias posdoctorales y sab\'aticas al extranjero para la consolidaci\'on de grupos de investigaci\'on\\
R.G. acknowledges support from the Centre National des Estudes Spatiales.\\
R.M. acknowledges support from CONICYT CATA-BASAL and FONDECYT through grant 3130750.\\
G.F., V.M., M.L, and R.C. acknowledge support from ECOS-CONICYT C12U02.\\
We acknowledge support from Programme National de Cosmologie (PNCG)\\
Based on observations obtained with MegaPrime/MegaCam, a joint project of CFHT and CEA/DAPNIA, at the Canada-France-Hawaii Telescope (CFHT) which is operated by the National Research Council (NRC) of Canada, the Institut National des Science de l'Univers of the Centre National de la Recherche Scientifique (CNRS) of France, and the University of Hawaii. This work is based in part on data products produced at TERAPIX and the Canadian Astronomy Data Centre as part of the Canada-France-Hawaii Tele- scope Legacy Survey, a collaborative project of NRC and CNRS.
\end{acknowledgements}

%

\bibliography{references}

\clearpage

\begin{table}[!ht]
\begin{center}
\label{table:results}
\begin{threeparttable}
\caption{Results of the weak lensing and optical analysis of the best SARCS candidates.}
\begin{tabular}{l c c c c c c c c c c c c  c}
\hline\hline\noalign{\smallskip}
Name & $z_{spec}$ & $z_{phot}$ &  $D_{ls}/D_{s}$ & $z_{eff}$ & $\sigma_{SIS}$ & $R_{A}$ & $R_{E}\,(z_{s})$ & $N_{0.5Mpc}$ & $L_{0.5Mpc}$ & $N_{1Mpc}$ & $L_{1Mpc}$\\
 & - & - & - & - & (km/s) & (arcsec) & (arcsec) (-)& - & $(10^{12}\mathrm{L_{\odot}})$ & - & $(10^{12}\mathrm{L_{\odot}})$\\
\noalign{\smallskip}\hline\noalign{\smallskip}
SA1 & - & 0.46 & 0.35 & 0.77 & $308_{-65}^{+150}$ & 2.2 & $1.6_{-0.9}^{+2.5}\,(1.42_{-0.57}^{+0.77})$ & $8\pm1$ & $0.52\pm0.07$ &  $13\pm3$ & $0.85\pm0.16$\\[3pt]
SA2 & - & 0.48 & 0.31 & 0.75 & $758_{-153}^{+88}$ & 5.0 & $9.4_{-5.4}^{+4.3}\,(1.43_{-0.56}^{+0.77})$ & $22\pm2$ & $1.58\pm0.13$ &  $40\pm4$ & $2.72\pm0.21$\\[3pt]
SA6 & - & 0.58 & 0.23 & 0.80 & $914_{-123}^{+89}$ & 5.0 & $12.0_{-6.5}^{+5.6}\,(1.45_{-0.54}^{+0.76})$ & $20\pm1$ & $1.61\pm0.08$ &  $26\pm2$ & $1.92\pm0.12$\\[3pt]
SA8 & - & 0.33 & 0.47 & 0.69 & $654_{-96}^{+61}$ & 10.8 & $8.5_{-3.7}^{+2.7}\,(1.40_{-0.59}^{+0.78})$ & $32\pm2$ & $1.63\pm0.16$ &  $60\pm6$ & $2.85\pm0.44$\\[3pt]
SA9 & - & 0.62 & 0.21 & 0.83 & $543_{-191}^{+139}$ & 3.3 & $4.0_{-3.0}^{+3.8}\,(1.47_{-0.53}^{+0.76})$ & $6\pm1$ & $0.78\pm0.06$ &  $12\pm2$ & $1.20\pm0.12$\\[3pt]
SA10 & - & 0.49 & 0.30 & 0.75 & $856_{-100}^{+114}$ & 3.2 & $11.9_{-5.7}^{+6.0}\,(1.43_{-0.56}^{+0.77})$ & $29\pm2$ & $2.37\pm0.17$  & $56\pm5$ & $3.65\pm0.36$\\[3pt]
SA11 & - & 0.62 & 0.20 & 0.81 & $458_{-175}^{+134}$ & 4.3 & $2.9_{-2.2}^{+3.1}\,(1.47_{-0.53}^{+0.76})$ & $10\pm1$ & $0.84\pm0.06$  & $12\pm1$ & $0.94\pm0.06$\\[3pt]
SA12 & - & 0.74 & 0.15 & 0.91 & $464_{-138}^{+252}$ & 3.4 & $2.5_{-1.9}^{+5.3}\,(1.51_{-0.50}^{+0.74})$ & $10\pm1$ & $0.66\pm0.10$  & $16\pm3$ & $0.85\pm0.15$\\[3pt]
SA13 & - & 0.29 & 0.52 & 0.67 & $393_{-100}^{+88}$ & 3.5 & $3.2_{-1.8}^{+2.0}\,(1.39_{-0.59}^{+0.78})$ & $19\pm2$ & $1.09\pm0.13$  & $45\pm5$ & $1.67\pm0.22$\\[3pt]
SA15 & 0.44$^{e}$ & 0.43 & 0.35 & 0.71 & $534_{-152}^{+129}$ & 3.9 & $5.0_{-3.2}^{+3.9}\,(1.42_{-0.57}^{+0.77})$ & $8\pm0$ & $0.69\pm0.03$ & $10\pm1$ & $0.78\pm0.06$\\[3pt]
SA18 & - & 0.38 & 0.41 & 0.70 & $506_{-152}^{+106}$ & 2.0 & $4.8_{-3.0}^{+3.1}\,(1.41_{-0.58}^{+0.78})$ & $8\pm1$ & $0.96\pm0.09$  & $13\pm2$ & $1.33\pm0.17$\\[3pt]
SA22 & 0.44$^{a}$ & 0.48 & 0.36 & 0.74 & $638_{-152}^{+101}$ & 7.1 & $7.0_{-4.2}^{+3.8}\,(1.42_{-0.57}^{+0.77})$ & $31\pm2$ & $1.73\pm0.10$ & $42\pm4$ & $2.21\pm0.19$\\[3pt]
SA23 & 0.61$^{a}$ & 0.74 & 0.21 & 0.81 & $713_{-192}^{+163}$ & 1.9 & $7.0_{-4.8}^{+6.1}\,(1.46_{-0.54}^{+0.76})$ & $7\pm1$ & $0.59\pm0.05$ & $10\pm2$ & $0.75\pm0.07$\\[3pt]
SA26 & - & 0.69 & 0.17 & 0.88 & $853_{-153}^{+145}$ & 16.4 & $9.0_{-5.7}^{+6.8}\,(1.49_{-0.52}^{+0.75})$ & $31\pm2$ & $2.67\pm0.15$ & $67\pm4$ & $5.60\pm0.31$\\[3pt]
SA27 & - & 0.18 & 0.66 & 0.59 & $643_{-78}^{+74}$ & 2.8 & $9.8_{-3.1}^{+3.0}\,(1.39_{-0.60}^{+0.78})$ & $12\pm1$ & $0.71\pm0.07$ & $24\pm4$ & $1.36\pm0.20$\\[3pt]
SA28 & - & 0.86 & 0.13 & 1.04 & $748_{-259}^{+179}$ & 2.4 & $5.5_{-4.4}^{+6.1}\,(1.57_{-0.47}^{+0.72})$ & $8\pm1$ & $1.19\pm0.13$ & $23\pm4$ & $1.95\pm0.32$\\[3pt]
SA29 & - & 0.72 & 0.17 & 0.92 & $394_{-115}^{+209}$ & 2.4 & $1.9_{-1.4}^{+3.8}\,(1.51_{-0.51}^{+0.74})$ & $11\pm1$ & $1.38\pm0.10$ & $21\pm4$ & $1.78\pm0.23$\\[3pt]
SA30 & 0.43$^{e}$ & 0.45 & 0.37 & 0.74 & $509_{-159}^{+118}$ & 5.6 & $4.5_{-3.0}^{+3.4}\,(1.42_{-0.57}^{+0.77})$ & $9\pm1$ & $0.47\pm0.05$ & $16\pm3$ & $0.71\pm0.16$\\[3pt]
SA31 & - & 0.27 & 0.55 & 0.66 & $593_{-116}^{+73}$ & 3.2 & $7.5_{-3.5}^{+2.7}\,(1.39_{-0.59}^{+0.78})$ & $3\pm1$ & $0.55\pm0.04$ & $6\pm1$ & $0.65\pm0.08$\\[3pt]
SA33 & 0.64$^{a}$ & 0.42 & 0.22 & 0.87 & $380_{-104}^{+173}$ & 2.4 & $1.9_{-1.3}^{+3.2}\,(1.47_{-0.53}^{+0.75})$ & $14\pm1$ & $0.76\pm0.05$ & $28\pm3$ & $1.33\pm0.15$\\[3pt]
SA36 & - & 0.35 & 0.45 & 0.70 & $724_{-107}^{+65}$ & 3.0 & $10.1_{-4.5}^{+3.3}\,(1.40_{-0.58}^{+0.78})$ & $23\pm2$ & $2.6\pm0.18$ & $30\pm3$ & $3.18\pm0.28$\\[3pt]
SA37 & - & 0.79 & 0.14 & 0.96 & $641_{-203}^{+214}$ & 2.2 & $4.5_{-3.4}^{+6.1}\,(1.54_{-0.49}^{+0.73})$ & $6\pm1$ & $0.66\pm0.08$ & $25\pm4$ & $2.66\pm0.31$\\[3pt]
SA39 & 0.61$^{a}$ & 0.72 & 0.20 & 0.80 & $655_{-233}^{+124}$ & 5.2 & $5.9_{-4.5}^{+4.4}\,(1.46_{-0.54}^{+0.76})$ & $5\pm0$ & $0.58\pm0.05$ & $8\pm2$ & $0.64\pm0.07$\\[3pt]
SA42 & - & 0.98 & 0.08 & 1.09 & $1049_{-315}^{+241}$ & 2.6 & $9.3_{-7.3}^{+11.0}\,(1.64_{-0.45}^{+0.71})$ & $8\pm1$ & $2.04\pm0.26$ & $16\pm4$ & $3.26\pm0.57$\\[3pt]
SA45 & - & 0.68 & 0.23 & 0.95 & $634_{-198}^{+116}$ & 3.5 & $5.1_{-3.7}^{+4.0}\,(1.49_{-0.52}^{+0.75})$ & $7\pm1$ & $0.51\pm0.08$ & $9\pm2^{(*)}$ & $0.69\pm0.14^{(*)}$\\[3pt]
SA47 & - & 0.80 & 0.19 & 1.06 & $390_{-115}^{+154}$ & 1.9 & $1.6_{-1.2}^{+2.6}\,(1.54_{-0.49}^{+0.73})$ & $4\pm1^{(*)}$ & $0.33\pm0.06^{(*)}$ & $8\pm1^{(*)}$ & $0.45\pm0.09^{(*)}$\\[3pt]
SA48 & 0.24$^{b}$ & 0.52 & 0.62 & 0.70 & $481_{-80}^{+60}$ & 2.8 & $5.1_{-2.1}^{+1.8}\,(1.39_{-0.60}^{+0.78})$ &  $2\pm0$ & $0.38\pm0.03$ & $13\pm3$ & $0.74\pm0.18$\\[3pt]
SA49 & - & 0.62 & 0.28 & 0.94 & $312_{-72}^{+125}$ & 4.3 & $1.3_{-0.9}^{+1.9}\,(1.47_{-0.53}^{+0.76})$ & $5\pm1$ & $0.34\pm0.05$ & $11\pm2$ & $0.60\pm0.11$\\[3pt]
SA50 & 0.51$^{c}$ & 0.54 & 0.29 & 0.77 & $540_{-172}^{+130}$ & 5.8 & $4.6_{-3.2}^{+3.8}\,(1.43_{-0.56}^{+0.77})$ & $18\pm1$ & $0.90\pm0.09$ & $24\pm2$ & $1.32\pm0.15$\\[3pt]
SA52 & - & 0.53 & 0.26 & 0.77 & $391_{-135}^{+136}$ & 2.1 & $2.3_{-1.7}^{+2.7}\,(1.44_{-0.55}^{+0.76})$ & $7\pm1$ & $0.49\pm0.04$ & $11\pm2$ & $0.58\pm0.08$\\[3pt]
SA53 & - & 0.55 & 0.27 & 0.81 & $389_{-130}^{+187}$ & 3.9 & $2.3_{-1.6}^{+3.7}\,(1.45_{-0.55}^{+0.76})$ & $4\pm1$ & $0.34\pm0.04$ & $12\pm3$ & $0.63\pm0.18$\\[3pt]
SA54 & - & 0.45 & 0.35 & 0.76 & $793_{-96}^{+72}$ & 6.3 & $10.7_{-5.0}^{+4.0}\,(1.42_{-0.57}^{+0.77})$ & $29\pm2$ & $2.10\pm0.18$ & $61\pm5$ & $3.73\pm0.38$\\[3pt]
SA55 & - & 0.38 & 0.42 & 0.72 & $701_{-109}^{+69}$ & 2.6 & $9.1_{-4.3}^{+3.3}\,(1.41_{-0.58}^{+0.78})$ & $23\pm2$ & $1.49\pm0.20$ & $41\pm5$ & $2.58\pm0.47$\\[3pt]
SA58 & - & 0.46 & 0.34 & 0.75 & $632_{-187}^{+92}$ & 2.6 & $6.7_{-4.4}^{+3.5}\,(1.42_{-0.57}^{+0.77})$ & $17\pm1$ & $0.92\pm0.06$ & $31\pm3$ & $1.61\pm0.12$\\[3pt]
SA59 & - & 0.79 & 0.11 & 0.92 & $838_{-322}^{+251}$ & 1.9 & $7.6_{-6.2}^{+9.6}\,(1.54_{-0.49}^{+0.73})$ & $4\pm1$ & $1.08\pm0.05$ & $14\pm3$ & $2.15\pm0.28$\\[3pt]
SA61 & - & 0.51 & 0.29 & 0.77 & $677_{-133}^{+108}$ & 7.4 & $7.2_{-4.2}^{+4.3}\,(1.43_{-0.56}^{+0.77})$ & $24\pm2$ & $2.27\pm0.16$ & $46\pm5$ & $3.30\pm0.35$\\[3pt]
SA63 & - & 0.48 & 0.34 & 0.79 & $561_{-155}^{+116}$ & 5.0 & $5.2_{-3.3}^{+3.6}\,(1.43_{-0.56}^{+0.77})$ & $12\pm1$ & $0.84\pm0.08$ & $20\pm3$ & $1.51\pm0.16$\\[3pt]
SA66 & 0.35$^{a}$ & 0.48 & 0.43 & 0.67 & $644_{-102}^{+69}$ & 4.8 & $8.0_{-3.7}^{+2.9}\,(1.40_{-0.58}^{+0.78})$ & $32\pm2$ & $1.90\pm0.09$ & $44\pm3$ & $2.42\pm0.17$\\[3pt]
SA67 & - & 0.45 & 0.34 & 0.73 & $591_{-150}^{+131}$ & 2.1 & $6.0_{-3.7}^{+4.3}\,(1.42_{-0.57}^{+0.77})$ & $8\pm1$ & $0.44\pm0.05$ & $9\pm1$ & $0.62\pm0.09$\\[3pt]
SA68 & - & 0.42 & 0.38 & 0.74 & $549_{-167}^{+100}$ & 2.8 & $5.3_{-3.5}^{+3.2}\,(1.41_{-0.57}^{+0.77})$ & $6\pm1$ & $0.59\pm0.10$ & $11\pm3$ & $1.41\pm0.15$\\[3pt]
SA70 & - & 0.29 & 0.51 & 0.66 & $438_{-125}^{+111}$ & 3.9 & $4.0_{-2.3}^{+2.8}\,(1.39_{-0.59}^{+0.78})$ & $9\pm1$ & $0.47\pm0.07$ & $19\pm2$ & $0.97\pm0.11$\\[3pt]
SA72 & 0.64$^{a}$ & 0.70 & 0.19 & 0.83 & $466_{-160}^{+150}$ & 4.5 & $2.9_{-2.2}^{+3.4}\,(1.47_{-0.53}^{+0.75})$ & $15\pm2$ & $1.31\pm0.15$ & $15\pm2$ & $1.44\pm0.16$\\[3pt]
SA74 & - & 0.36 & 0.45 & 0.72 & $672_{-94}^{+89}$ & 3.2 & $8.6_{-3.8}^{+3.7}\,(1.40_{-0.58}^{+0.78})$ & $21\pm2$ & $1.34\pm0.13$ & $44\pm4$ & $2.83\pm0.31$\\[3pt]
SA78 & - & 0.74 & 0.21 & 1.00 & $543_{-157}^{+96}$ & 3.2 & $3.4_{-2.5}^{+2.8}\,(1.51_{-0.50}^{+0.74})$ & $30\pm3$ & $2.26\pm0.18$ & $56\pm6$ & $3.78\pm0.35$\\[3pt]
SA79 & - & 0.76 & 0.21 & 1.03 & $340_{-60}^{+130}$ & 3.5 & $1.3_{-0.8}^{+2.0}\,(1.52_{-0.50}^{+0.74})$ & $12\pm2$ & $0.95\pm0.13$ &  $55\pm7$ & $4.03\pm0.44$\\[3pt]
SA80 & - & 1.00 & 0.11 & 1.17 & $536_{-193}^{+149}$ & 2.4 & $2.4_{-1.9}^{+3.2}\,(1.65_{-0.44}^{+0.70})$ & $20\pm1$ & $1.39\pm0.14$ & $35\pm5$ & $2.78\pm0.44$\\[3pt]
SA84 & - & 0.77 & 0.21 & 1.05 & $513_{-169}^{+86}$ & 1.9 & $2.9_{-2.3}^{+2.4}\,(1.53_{-0.50}^{+0.74})$ & $6\pm1$ & $0.28\pm0.08$ & $11\pm2$ & $0.74\pm0.14$\\[3pt]
SA86 & - & 0.46 & 0.35 & 0.77 & $600_{-151}^{+82}$ & 3.7 & $6.1_{-3.7}^{+3.0}\,(1.42_{-0.57}^{+0.77})$ & $16\pm1$ & $1.49\pm0.11$ & $41\pm3$ & $2.88\pm0.25$\\[3pt]
SA87 & - & 0.54 & 0.25 & 0.77 & $425_{-124}^{+192}$ & 3.5 & $2.7_{-1.9}^{+4.2}\,(1.44_{-0.55}^{+0.76})$ & $14\pm1$ & $1.16\pm0.12$ & $29\pm3$ & $2.35\pm0.28$\\[3pt]
\end{tabular}
  \end{threeparttable}
  \end{center}
\end{table}

\clearpage

\begin{table}[!ht]
\begin{center} 
\begin{threeparttable}
\begin{tabular}{l c c c c c c c c c c c c c}
\hline\hline\noalign{\smallskip}
Name & $z_{spec}$ & $z_{phot}$ &  $D_{ls}/D_{s}$ & $z_{eff}$ & $\sigma_{SIS}$  & $R_{A}$ & $R_{E}\,(z_{s})$ & $N_{0.5Mpc}$ & $L_{0.5Mpc}$ & $N_{1Mpc}$ & $L_{1Mpc}$\\
 & - & - & - & - & (km/s) & (arcsec) & (arcsec) & - & $(10^{12}\mathrm{L_{\odot}})$ & - & $(10^{12}\mathrm{L_{\odot}}$)\\
\noalign{\smallskip}\hline\noalign{\smallskip}
SA89 & - & 0.42 & 0.34 & 0.68 & $676_{-163}^{+96}$ & 3.7 & $8.1_{-4.8}^{+4.0}\,(1.41_{-0.57}^{+0.77})$ & $9\pm1$ & $0.95\pm0.13$ & $37\pm5$ & $2.93\pm0.40$\\[3pt]
SA90 & - & 0.53 & 0.26 & 0.77 & $1015_{-79}^{+70}$ & 3.7 & $15.8_{-7.2}^{+5.8}\,(1.44_{-0.55}^{+0.76})$ & $20\pm2$ & $1.52\pm0.12$ & $45\pm5$ & $2.82\pm0.27$\\[3pt]
SA91 & - & 0.56 & 0.24 & 0.78 & $360_{-87}^{+195}$ & 3.0 & $1.9_{-1.2}^{+3.6}\,(1.45_{-0.55}^{+0.76})$ & $13\pm1$ & $1.03\pm0.09$ & $22\pm3$ & $1.64\pm0.19$\\[3pt]
SA92 & - & 0.50 & 0.31 &  0.78 & $595_{-164}^{+105}$ & 2.8 & $5.7_{-3.7}^{+3.6}\,(1.43_{-0.56}^{+0.77})$ & $8\pm1$ & $0.44\pm0.06$ & $11\pm1^{(*)}$ & $0.53\pm0.08^{(*)}$\\[3pt]
SA94 & - & 0.51 & 0.29 & 0.76 & $653_{-196}^{+100}$ & 0.0 & $6.7_{-4.6}^{+3.9}\,(1.43_{-0.56}^{+0.77})$ & $16\pm1$ & $1.01\pm0.12$ & $27\pm5$ & $1.99\pm0.32$\\[3pt]
SA95 & - & 0.49 & 0.28 & 0.73 & $392_{-112}^{+197}$ & 2.2 & $2.5_{-1.6}^{+4.1}\,(1.43_{-0.56}^{+0.77})$ & $3\pm1$ & $0.70\pm0.07$ & $7\pm2$ & $0.81\pm0.13$\\[3pt]
SA96 & - & 0.39 & 0.37 & 0.67 & $449_{-138}^{+123}$ & 2.8 & $3.7_{-2.4}^{+3.1}\,(1.41_{-0.58}^{+0.78})$ & $13\pm2$ & $0.80\pm0.13$ & $40\pm5$ & $2.74\pm0.42$\\[3pt]
SA97 & 0.42$^{c}$ & 0.48 & 0.34 & 0.67 & $384_{-109}^{+162}$ & 8.0 & $2.6_{-1.7}^{+3.4}\,(1.41_{-0.57}^{+0.77})$ & $15\pm1$ & $0.89\pm0.09$ & $28\pm4$ & $1.09\pm0.23$\\[3pt]
SA98 & - & 0.52 & 0.28 & 0.77 & $932_{-133}^{+107}$ & 18.4 & $13.5_{-7.1}^{+6.5}\,(1.44_{-0.56}^{+0.77})$ & $29\pm2$ & $2.08\pm0.21$ & $56\pm3$ & $3.37\pm0.46$\\[3pt]
SA99 & - & 0.32 & 0.48 & 0.67 & $395_{-118}^{+103}$ & 2.4 & $3.1_{-1.9}^{+2.3}\,(1.40_{-0.59}^{+0.78})$ & $4\pm1$ & $0.46\pm0.10$ & $7\pm2$ & $0.54\pm0.12$\\[3pt]
SA100 & - & 0.63 & 0.19 & 0.82 & $969_{-130}^{+100}$ & 14.7 & $12.6_{-7.1}^{+6.5}\,(1.47_{-0.53}^{+0.75})$ & $30\pm2$ & $3.00\pm0.16$ & $71\pm4$ & $5.49\pm0.35$\\[3pt]
SA101 & - & 0.87 & 0.10 & 0.99 & $902_{-268}^{+192}$ & 3.5 & $7.9_{-6.1}^{+8.2}\,(1.58_{-0.47}^{+0.72})$ & $5\pm1$ & $2.36\pm0.24$ & $18\pm5$ & $3.36\pm0.54$\\[3pt]
SA102 & - & 0.69 & 0.16 & 0.86 & $1028_{-272}^{+140}$ & 9.9 & $13.1_{-9.2}^{+8.5}\,(1.49_{-0.52}^{+0.75})$ & $32\pm2$ & $2.03\pm0.13$ & $58\pm4$ & $3.40\pm0.27$\\[3pt]
SA103 & - & 0.47 & 0.28 & 0.70 & $450_{-134}^{+164}$ & 4.1 & $3.4_{-2.2}^{+3.9}\,(1.42_{-0.57}^{+0.77})$ & $12\pm1$ & $0.86\pm0.08$ & $25\pm4$ & $1.46\pm0.22$\\[3pt]
SA106 & - & 0.74 & 0.15 & 0.91 & $490_{-176}^{+185}$ & 1.9 & $2.8_{-2.2}^{+4.1}\,(1.51_{-0.50}^{+0.74})$ & $7\pm1$ & $0.50\pm0.11$ & $16\pm3$ & $1.14\pm0.21$\\[3pt]
SA108 & - & 0.86 & 0.10 & 0.99 & $1008_{-345}^{+213}$ & 4.5 & $10.1_{-8.0}^{+10.2}\,(1.57_{-0.47}^{+0.72})$ & $25\pm1$ & $2.01\pm0.20$ & $41\pm4$ & $3.18\pm0.39$\\[3pt]
SA109 & - & 0.39 & 0.41 & 0.72 & $394_{-100}^{+183}$ & 3.2 & $2.9_{-1.7}^{+4.1}\,(1.41_{-0.58}^{+0.78})$  & $3\pm1$ & $0.47\pm0.11$ & $8\pm3^{(*)}$ & $0.90\pm0.18^{(*)}$\\[3pt]
SA110 & - & 0.18 & 0.66 & 0.58 & $329_{-80}^{+104}$ & 4.1 & $2.6_{-1.3}^{+2.1}\,(1.39_{-0.60}^{+0.78})$ & $4\pm1$ & $0.42\pm0.07$ & $16\pm4$ & $2.09\pm0.24$\\[3pt]
SA111 & - & 0.52 & 0.29 & 0.79 & $889_{-88}^{+67}$ & 5.0 & $12.3_{-5.8}^{+4.6}\,(1.44_{-0.56}^{+0.77})$ & $28\pm2$ & $2.21\pm0.13$ & $49\pm4$ & $3.73\pm0.26$\\[3pt]
SA112 & 0.50$^{d}$ & 0.55 & 0.29 & 0.75 & $626_{-178}^{+112}$ & 4.3 & $6.3_{-4.1}^{+4.0}\,(1.43_{-0.56}^{+0.77})$ & $20\pm1$ & $0.95\pm0.08$ & $43\pm4$ & $1.85\pm0.19$\\[3pt]
SA113 & 0.67$^{d}$ & 0.71 & 0.17 & 0.85 & $745_{-210}^{+139}$ & 3.0 & $7.1_{-5.0}^{+5.6}\,(1.49_{-0.52}^{+0.75})$ & $16\pm2$ & $1.25\pm0.12$ & $32\pm3$ & $2.21\pm0.23$\\[3pt]
SA114 & - & 0.83 & 0.11 & 0.97 & $809_{-300}^{+168}$ & 3.5 & $6.7_{-5.4}^{+6.6}\,(1.56_{-0.48}^{+0.73})$ &  $14\pm1$ & $1.30\pm0.14$ & $48\pm4$ & $3.72\pm0.37$\\[3pt]
SA116 & - & 0.57 & 0.24 & 0.80 & $706_{-216}^{+126}$ & 4.1 & $7.3_{-5.1}^{+5.0}\,(1.45_{-0.54}^{+0.76})$ & $8\pm1$ & $0.54\pm0.05$ & $15\pm3$ & $0.81\pm0.14$\\[3pt]
SA117 & - & 0.43 & 0.37 & 0.74 & $697_{-111}^{+92}$ & 7.3 & $8.5_{-4.2}^{+4.0}\,(1.42_{-0.57}^{+0.77})$ & $26\pm3$ & $1.51\pm0.24$ & $45\pm7$ & $1.98\pm0.41$\\[3pt]
SA120 & - & 0.46 & 0.24 & 0.76 & $556_{-172}^{+91}$ & 2.1 & $5.2_{-3.5}^{+3.0}\,(1.42_{-0.57}^{+0.77})$ & $6\pm1$ & $0.34\pm0.04$ & $9\pm2$ & $0.54\pm0.13$\\[3pt]
SA121 & - & 0.62 & 0.12 & 0.84 & $443_{-137}^{+170}$ & 3.7 & $2.7_{-1.9}^{+3.7}\,(1.47_{-0.53}^{+0.76})$ & $10\pm2$ & $1.03\pm0.13$ & $16\pm4$ & $1.76\pm0.26$\\[3pt]
SA123 & - & 1.00 & 0.07 & 1.11 & $894_{-288}^{+267}$ & 4.8 & $6.6_{-5.3}^{+9.5}\,(1.65_{-0.44}^{+0.70})$ & $18\pm2$ & $1.73\pm0.16$ & $37\pm4$ & $3.70\pm0.42$\\[3pt]
SA124 & - & 0.83 & 0.20 & 1.11 & $545_{-164}^{+102}$ & 7.4 & $3.1_{-2.3}^{+2.8}\,(1.56_{-0.48}^{+0.73})$ & $12\pm1$ & $0.56\pm0.07$ & $20\pm2$ & $0.93\pm0.14$\\[3pt]
SA125 & - & 0.74 & 0.15 & 0.91 & $695_{-235}^{+176}$ & 0.0 & $5.6_{-4.3}^{+5.9}\,(1.51_{-0.50}^{+0.74})$ & $24\pm2$ & $2.75\pm0.22$ & $33\pm4$ & $3.66\pm0.34$\\[3pt]
SA127 & 0.33$^{c}$ & 0.33 & 0.47 & 0.70 & $645_{-123}^{+68}$ & 4.7 & $8.2_{-4.0}^{+2.9}\,(1.40_{-0.59}^{+0.78})$ & $5\pm1$ & $0.56\pm0.05$ & $11\pm3$ & $0.74\pm0.12$\\[3pt]
SA0 & 0.38$^{a}$ & - & 0.41 & 0.71 & $638_{-135}^{+85}$ & 6.1 & $7.6_{-4.1}^{+3.4}\,(1.41_{-0.58}^{+0.78})$ & $11\pm1$ & $0.84\pm0.04$ & $21\pm3$ & $1.26\pm0.12$\\[3pt]
\hline
\end{tabular}
    \begin{tablenotes}
      \small
      \item (*): groups that are closer to the edge of the field of view than the aperture in which richness and luminosity are estimated.
      \item (a): \cite{munoz13}. (b): \cite{ruff11}. (c): \cite{sl2s}. (d): \cite{thanjavur10}. (e): Motta et al. (in prep.).
      \item Columns: (1) SARCS name. (2) Spectroscopic redshift. (3) Photometric redshift from \cite{coupon09}. (4) Average geometrical factor. (5) Effective redshift derived from the average geometrical factor. (6) SIS velocity dispersion derived from the shear profile (quoted errors are the statistical uncertainties from the shear profile fitting). (7) Arc radius (from \citealt{more12}). (8) Einstein radius from $\sigma_{SIS}$ and the source redshift in parenthesis (error bars account both for errors on $\sigma_{SIS}$ and the PDF of $z_{s}$). (9-10) Optical richness and luminosity derived from the bright red galaxies. (11-12) Same as (9-10) but using an aperture of 1 Mpc. 
    \end{tablenotes}
  \end{threeparttable}
  \end{center}
\end{table}

\clearpage

\begin{center}
\section*{Appendix}
\end{center}

\begin{figure*}
\center
\includegraphics[width=15cm, angle=0]{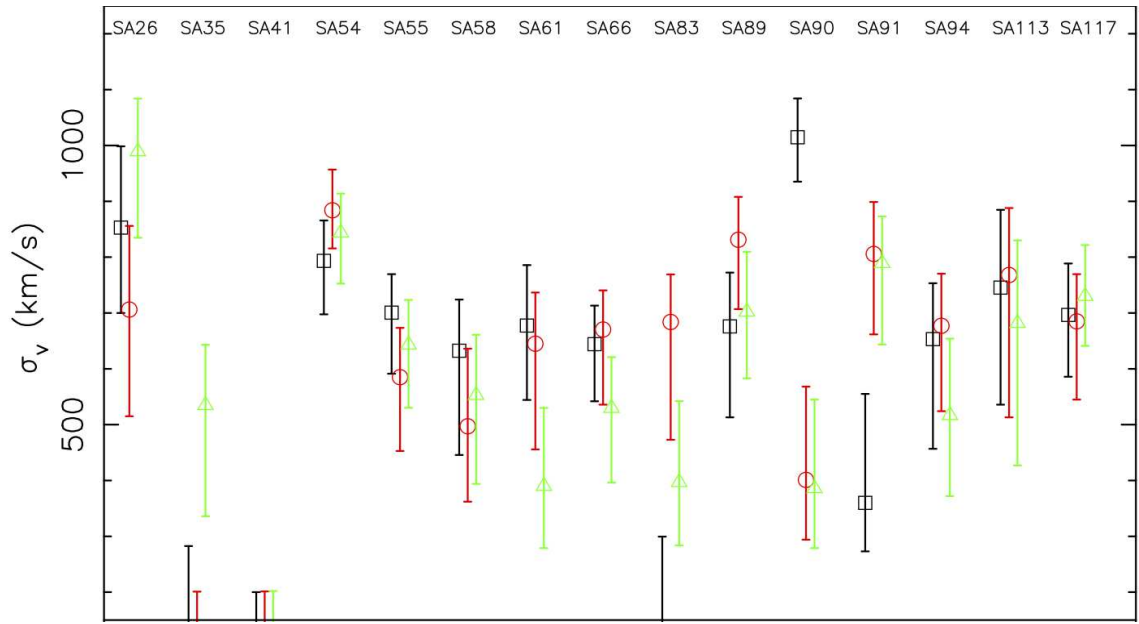}
\caption{SIS model on the multimodal SARCS candidates. The centre chosen for the shear profile is located either on the strong lensing system (black open squares), at the middle of the 2 luminosity over-densities  (red open circles) or on the luminosity peak not associated to the strong lensing features (green open triangles). The measured dispersion at the three different positionings can be used as an indicator of the dominance of the strong lens within its group.}
\label{fig:bimod} 
\end{figure*}

In section 3.2 we briefly discuss the issue of choosing the most likely centre of the halo mass distribution. From the optical luminosity maps, we have shown that several SARCS objects present a complex morphology. Assuming that light traces mass at the group and cluster-scale (e.g. \citealt{bahcall00}), these substructures in the distribution of galaxies might be associated to massive sub-haloes. As the weak lensing estimator is sensitive to all the mass components where the signal is measured, a fit using a single halo will be affected by all the present substructures. However, the question of the centre remains as a source of uncertainties: substantial miscentering can lead to weak lensing masses underestimated up to 30\% (e.g. \citealt{george12}).\\
Here we explore the effect of changing the position of the centre used to construct the shear profile. We limit the analysis to the SARCS groups with a multimodal structure in their luminosity maps within a 0.5 Mpc radius from the centre of the strong lensing system. For group-scale haloes, typical values of the Virial radius are $\sim1$ Mpc, so within a 0.5 Mpc radius we expect to rather observe substructures than two distinct haloes. Hence, fitting shear profiles with a single component remains valid. So we changed the position of the centre of the shear profile for these groups, simply positioning it either between the two optical over-densities or on the second observed peak, i.e. not associated to the strong lensing system.\\
The results we obtained are presented Figure \ref{fig:bimod}. For two groups, SA35 and SA83, we managed to obtain better constraints than in the initial configuration. It suggests that the strong lensing system is not exactly at the mass centre but rather associated to a satellite galaxy. For SA90, we observe a strong change according to the centre position, with a $\sigma_{SIS}$ much higher when using the strong lensing system as the centre of the shear profile. For this group, the brightest galaxy is also at the centre of the strong lensing system, which seems to indicate that the mass associated to the second luminosity peak is negligible compared to the main halo. SA91 presents the opposite behavior, with a velocity dispersion higher when the centre of the shear profile is moved towards the second luminosity peak. As for SA35 and SA83, we suppose that the strong lensing system is associated with a satellite galaxy.\\
In the remaining cases, we only observe slight variations with compatible velocity dispersions within their $1\sigma$ error bar, which makes the interpretation of the results speculative. For groups that have the highest $\sigma_{SIS}$ when the centre of the shear profile is taken between the two luminosity over-densities as SA89, one can explain such a variation by the presence of two clumps of galaxies evolving in a single dark matter halo which mass centre is located at the middle of the galaxy distribution. For instance, SA66 was studied in more details by \cite{limousin10} with a strong lensing modeling that requires a substantial external shear, and by \cite{munoz13}, with a dynamical analysis that revealed the presence of two distinct populations of galaxies. In that case, the results suggest a merging event of two sub-haloes. Such a configuration is consistent with the weak lensing results as moving the centre of the shear profile in both directions from the global mass centre induces the observed lower velocity dispersions, with a larger decrease when going towards the second luminosity peak (not associated with the strong lensing system). It is therefore tempting to infer the same for the groups showing the same variation of $\sigma_{SIS}$. One can also think of two distinct dark matter haloes with similar masses that generate their own shear signal, and thus we observe the opposite variation with lower values of $\sigma_{SIS}$ when taking the centre of the profiles between the two luminosity over-densities (SA26, SA55). Another possible configuration would be a mass distribution dominated by a halo located on the strong lensing system, and in that case, the measured velocity dispersion decreases when moving away, such as for SA61.

Despite we observe statistically significant changes for some groups, given the weakness of the signal we measure on single objects, it remains difficult to probe the position of the actual mass centre via weak lensing and draw reliable conclusions from the small observed variations in the shear profile. Our first assumption of positioning the mass centre on the strong lensing system then remains on average a valid approximation. On specific cases, a deeper analysis combining a strong lensing modeling (external shear) and dynamical information (two distinct populations of galaxies) might however help to increase the precision of the mass determination for such complex configurations.

\end{document}